\documentclass[aps,prd,fleqn,twocolumn,amsmath,amssymb,floatfix]{revtex4}
\RequirePackage{graphicx}
\RequirePackage{flushend}
\RequirePackage[colorlinks,citecolor=blue,urlcolor=blue,linkcolor=blue]{hyperref}
\RequirePackage{xparse}
\usepackage{physics}

\newcommand{\rt}{r\hat{n},t}
\newcommand{\rtl}{r_L\hat{n},t_L}

\begin{document}
\title{An analytical approach to the CMB anisotropies in a Spatially Closed background}
\author         {Pedram Niazy}
\email          {pedram.niazi@modares.ac.ir}
\author         {Amir H. Abbassi}
\email          {ahabbasi@modares.ac.ir}

\affiliation    {Department of Physics, School of Science, Tarbiat Modares University. P.O. Box 14155-4838, Tehran, Iran}
\date{\today}

\begin{abstract}
The scalar mode temperature fluctuations of the cosmic microwave background has been derived in a spatially closed universe from two different methods. First, by following the photon trajectory after the last scattering and then from the Boltzmann equation in a closed background and the line of sight integral method. An \emph{analytic} expression for the temperature multipole coefficient has been extracted at the hydrodynamical limit, where we have considered some tolerable approximations. By considering a realistic set of cosmological parameters taken from a fit to data from Planck, the TT power spectrum in the scalar mode for the closed universe has been compared with numerical one by using the CAMB code and also latest observational data. The analytic result agrees with the numerical one on almost all scales. The peak positions are in very good agreement with numerical result while the peak heights agree with that to within $10\%$ due to the approximations have been considered for this derivation.\\
\end{abstract}

\maketitle

\section{Introduction}\label{501}
The cosmological parameters of the standard big bang model can be determined or considerably constrained by comparing the predictions of theoretical cosmological models with the data on the CMB by observation, such as WMAP and Planck.
The theoretical derivation of the spectra of the CMB temperature anisotropies and polarizations has been archived by sophisticated numerical calculation codes such as CAMB \cite{r1} and CMBFAST \cite{r2} that give the spectra $C_{XX',\ell}$ which involves several cosmological parameters. However, analytical studies give us a great insight into the problem for understanding how various underlying physical effects give rise to specific observational behavior. In particular, the analytical studies are helpful in revealing the explicit dependencies of the CMB spectra on cosmological parameters and possible degeneracies between them.\\
There are several works in the field that extracted an analytical expression for the $C_{XX',\ell}$ in a flat universe. In Refs. \cite{r3,r4,r5,r6,r7,r8,r9} you can find all analytical spectra by considering the tensor perturbation as a source. Refs. \cite{r10,r11} gave the analytic calculation of the scalar mode temperature power spectrum in Newtonian gauge while Refs. \cite{r12,r13} gave the scalar mode analytic power spectra in synchronous gauge. Ref. \cite{r14} also gave a unifying framework for all spectra in both tensor and scalar modes. However, the analysis is still incomplete by lacking analytic expressions for the closed and open geometry. viewing these, we are going to perform a detailed analytic calculation of scalar mode (in synchronous gauge) temperature power spectrum $C_{TT,\ell}^S$ in a spatially closed background. We will apply some of the results and techniques developed in the study of the cosmic microwave background anisotropies in a flat spatial geometry to the closed case and compare the consequences with that of numerical calculation and latest observational data. By applying a series of tolerable approximations that lead to a simple \emph{analytic} formula for the CMB power spectrum, we provide transparent information about the dependencies of the CMB spectra on cosmological parameters. We extract an analytic formula for the closed universe temperature fluctuations by imposing the effect of curvature into the Boltzmann equation for the photons and using the line of sight method without using any recursion relations which used by others for numerical calculations. We derive this expression from a more geometrical approach, by following the photon trajectory from the last scattering surface until now in a spatially closed background. We also calculate the multipole coefficient \emph{analytically}, in hydrodynamic limit and compare the result with numerical calculations and observational data.\\
In the following section, we give a brief overview of the perturbation theory in a spatially closed universe and its applications in the present paper. In section \ref{503}, we extract the temperature fluctuations by following the photon trajectory from the last scattering surface in the spatially closed background. In section \ref{504}, we introduce the Boltzmann equation for the photons in the spatially closed background and extract the temperature fluctuations by using the line of sight integral method. The approach presented here for temperature fluctuations can also be used for extracting the polarization multipoles $C_{TE,\ell}^S$ and $C_{EE,\ell}^S$  from the Boltzmann equation. In section \ref{505}, at first, we introduce a general formula for the temperature multipole coefficient in a closed background and then by considering that the evolution of cosmological perturbations is primarily hydrodynamics, among some other appropriate approximations, we extract an analytic formula for the temperature power spectrum $C_{TT,\ell}^S$\,. In section \ref{506}, we plot the TT power spectrum curve extracted in section \ref{505} using a realistic set of cosmological parameters and compare it with the numerical one by CAMB and also the curve from latest observational data (Planck 2015). Several interesting properties of CMB anisotropies are revealed in analytic expression along with the power spectrum dependence on cosmological parameters. We conclude the article by a brief review that remarks the main outcomes of this paper.

\section{The perturbation theory in a spatially closed background; a short review}\label{502}
The theory of the linear perturbations is an important part of the modern cosmology which explains CMB anisotropies and the origin of structure formation. There is enough references for this theory in a spatially flat universe and has been investigated for a spatially closed universe recently \cite{r15}.\\
The perturbed metric is:
\begin{equation}\label{1}
  g_{\mu\nu}=\overline{g}_{\mu\nu}+h_{\mu\nu}
\end{equation}
where $\overline{g}_{\mu\nu}$ and $h_{\mu\nu}$ are the unperturbed metric and the first order perturbation, respectively. Note that $\overline{g}_{\mu\nu}$ is the FLRW metric which in the comoving spherical polar coordinates can be written as
\begin{align*}
  \overline{g}_{00}=&-1\\
  \overline{g}_{rr}=&{\frac{a^2 (t)}{1-Kr^2}\qquad
  \overline{g}_{\theta\theta}=a^2(t)r^2\qquad
  \overline{g}_{\varphi\varphi}=a^2(t)r^2\sin^2\theta}
\end{align*}
Perturbation in the metric leads to perturbation in the Ricci and energy-momentum tensor. We can decompose the metric perturbation and energy-momentum tensors into the scalar, vector and tensor modes from their transformation properties under spatial rotations and derive the field equations accordingly \cite{r15}.\\
Decomposition into the scalar, vector and tensor modes of the metric perturbation and energy-momentum tensor would be as follows:
\begin{align*}
  h_{00} =& -E \\
  h_{i0} =& a\,(\nabla_i F+G_i)\\
  h_{ij} =& a^2\,(A\,\widetilde{g}_{ij}+H_{ij} B+\nabla_i C_j+\nabla_j C_i+D_{ij})\\
  \delta T_{00} =& -\overline{\rho}\,h_{00}+\delta \rho\\
  \delta T_{i0} =& \overline{p}\,h_{i0}-(\overline{\rho} + \overline{p} )(\nabla_i \delta u+\delta u_i^V)\\
  \delta T_{ij} =& \overline{p}\, h_{ij}+a^2 (\widetilde{g}_{ij}\delta p+H_{ij} \Pi^S+\nabla_i \Pi_j^V+\nabla_j \Pi_i^V+\Pi_{ij}^T).
\end{align*}
where $\nabla_i$ is the covariant derivative with respect to the spatial unperturbed metric $\widetilde{g}_{ij}(=a^{-2}\overline{g}_{ij})$ and $H_{ij}=\nabla_i \nabla_j$ is the \emph{covariant Hessian operator}. All the perturbations $A\,$, $B\,$, $E\,$, $F\,$, $C_i\,$, $G_i\,$ and $D_{ij}$ are functions of $\textbf{x}$ and $t$ which satisfy
\begin{align*}
  &\nabla^i C_i=\nabla^i G_i=0\\
  &\widetilde{g}^{ij}D_{ij}=0\quad \nabla^i D_{ij}=0\quad D_{ij}=D_{ji}
\end{align*}
On the other hand, all above perturbative quantities have been considered as random fields on $S^3(\alpha)$ (a 3-sphere of radius $\alpha$), because they are defined on a homogeneous and isotropic space \cite{r16,r17}. So they can be described by their Fourier transformation. There are many different Fourier transform convention, however here we are going to expand each mode of the perturbation fields in terms of the corresponding eigenfunctions of the \emph{Laplace-Beltrami} operator. This operator reduces to the ordinary Laplacian in a flat background. In pseudo-spherical coordinates with the line element
\begin{equation}\label{2}
  ds^2=\alpha^2\,(\textrm d\chi^2+\sin^2 \chi\, \textrm d\theta^2 +\sin^2 \chi\, \sin^2 \theta \,\textrm d\varphi^2  )
\end{equation}
one gets the following eigenvalues and eigenfunctions for the Laplace-Beltrami operator:
\begin{align*}
  &\nabla^2\Phi =  -k_n^2 \Phi \qquad \nabla^2=\widetilde{g}_{ij}H_{ij}=\widetilde{g}_{ij}\nabla_i\nabla_j\\
  &\Phi = \mathcal{Y}_{n\ell m} (\chi,\theta,\varphi)=\Pi_{n\ell} (\chi)\, Y_{\ell m} (\theta,\varphi)\\
  &k_n^2 =  \frac{n^2-1}{\alpha^2}\qquad n=1,2,\cdots
\end{align*}
where $\Pi_{n\ell}(\chi)$ is the hyperspherical Bessel function satisfying the following equation
\begin{align}\label{3}
  \dv[2]{\Pi_{n\ell}(\chi)}{\chi}+&2\cot\chi\dv{\Pi_{n\ell}(\chi)}{\chi}\nonumber\\
  +&\left[(n^2-1)-\frac{\ell(\ell+1)}{\sin^2\chi}\right]{\Pi_{n\ell}(\chi)}=0
\end{align}
In a flat background, the hyperspherical Bessel function reduces to the ordinary spherical Bessel function $j_\ell (\nu\chi)$.
Also we introduce the generalized wave number in closed space $q_n$ as
\begin{align*}
    q_n = \sqrt{k_n^2 + \frac{1}{\alpha^2}} = \frac{n}{\alpha}
\end{align*}
We can expand the scalar perturbative quantity $A(\textbf{x},t)$ in terms of \emph{Laplace-Beltrami} operator eigenfunctions as below:
\begin{equation}\label{4}
 A(\textbf{x},t)=\sum_{n\ell m}A_{n\ell m}(t)\,\mathcal{Y}_{n\ell m}(\chi,\theta,\varphi)
\end{equation}
This is the initial conditions that depend on the direction, not the perturbation itself, so a perturbation can be shown by a time-dependent normal mode $A_n (t)$ with an overall normalization factor $\alpha_{lm}$.
$A_{n\ell m}(t)$ just like $A(\textbf{x},t)$ is a scalar random field and one of the simplest statistics for it is the two-point covariant function denoted by $\langle A_{n\ell m} A^\ast_{n'\ell'm'}\rangle$. Here $\langle\quad\rangle$ means the ensemble average which equals the spatial average according to the \emph{ergodic} theorem.
The homogeneity and isotropy imply that
\begin{equation*}
  \langle\alpha_{\ell m}\,\alpha_{\ell' m'}^\ast\rangle=\delta_{\ell\ell'}\,\delta_{mm'}\qquad\langle A_n(t)\,A_{n'}^\ast(t)\rangle=A_n^2(t)\,\delta_{nn'}
\end{equation*}
so the two-point covariant function of $A_{n\ell m}(t)$ is
\begin{align*}
  \langle A_{n\ell m} A^\ast_{n'\ell'm'}\rangle
   &= \langle\alpha_{\ell m}\,A_n(t)\,\alpha_{\ell m}^\ast\,A_{n'}^\ast(t)\rangle\\
   &=\langle\alpha_{\ell m}\,\alpha_{\ell' m'}^\ast\rangle\langle A_n(t)\,A_{n'}^\ast(t)\rangle\\
   &=A_n^2(t)\,\delta_{nn'}\,\delta_{\ell\ell'}\,\delta_{mm'}
\end{align*}
and for any scalar random field $A$, we will have
\begin{equation}\label{5}
 A(\textbf{x},t)=\sum_{n\ell m}\alpha_{\ell m}\,A_n(t)\,\mathcal{Y}_{n\ell m}(\chi,\theta,\varphi)
\end{equation}
Also, we can decompose an arbitrary tensor random field using the scalar eigenvalues of the Laplace-Beltrami operator and their covariant derivatives as follows
\begin{align}\label{6}
  A^{ij}(\textbf{x},t)&=\sum_{n\ell m}\alpha_{\ell m}\bigg[\frac{1}{3}A_{nT}(t)\,\tilde{g}^{ij}\,\mathcal{Y}_{nlm}(\chi,\theta,\varphi)\nonumber\\
  &+A_{nTL}(t)\left(k_n^{-2}H^{ij}\mathcal{Y}_{nlm}(\chi,\theta,\varphi)\right) \bigg],
\end{align}
where $A_{nT}(t)$ and $A_{nTL}(t)$ are the trace and traceless parts of the tensor $A^{ij}$ respectively \cite{r18,r19}.

\section{Temperature fluctuations by Following the photon trajectory}\label{503}
In this section we derive the scalar mode CMB temperature fluctuations by following the photon trajectory from the last scattering surface in a spatially closed background.\\
We can write the perturbed metric in any specific gauge with $g_{i0}=0$ in the form
\begin{align}\label{7}
  g_{00}&=-\left(1+E(r\hat{n},t)\right)\qquad g_{i0}=g_{0i}=0\nonumber\\
  g_{rr}&=\left(\frac{a^2(t)}{1-Kr^2}+h_{rr}(r\hat{n},t)\right)
\end{align}
A light ray travelling toward the center of the coordinate system from the direction $\hat{n}$ will have a comoving radial coordinate $r$ related to $t$ by
\begin{align}\label{8}
    0=-\left(1+E(r\hat{n},t)\right)\,\textrm dt^2+\left(\frac{a^2(t)}{1-Kr^2}+h_{rr}(r\hat{n},t)\right)\,\textrm dr^2
\end{align}
which gives
\begin{align}\label{9}
  \frac{\textrm dr}{\textrm dt}&=-\left(\frac{\frac{a^2}{1-Kr^2}+h_{rr}}{1+E}\right)^{-\frac{1}{2}}\nonumber\\
  &\simeq-\frac{\sqrt{1-Kr^2}}{a}\left(1-\frac{h_{rr}(1-Kr^2)}{2a^2}+\frac{E}{2}\right)
\end{align}
Now we make the assumption that the transition of cosmic matter from opacity to transparency occurred suddenly at a time $t_L$ of last scattering. With this approximation, the first-order solution of Eq.(\ref{9}) in spatially closed background is
\begin{equation}\label{10}
  \sin^{-1}r(t)=\sin^{-1}s(t)+\int_{t_L}^{t}\frac{\textrm dt'}{a(t')}N(s(t')\hat{n},t')
\end{equation}
where
\begin{equation*}
  N(\textbf{x},t)=\frac{1}{2}\left[\frac{h_{rr}(\textbf{x},t)}{a^2(t)}(1-r^2)-E(\textbf{x},t)\right]
\end{equation*}
and $\sin^{-1}s(t)$ is the zeroth order solutions for the radial coordinate which has the value $\sin^{-1}r_L$ at $t=t_L$:
\begin{equation}\label{11}
  \sin^{-1}s(t)=\sin^{-1}r_L-\int_{t_L}^{t}\frac{\textrm dt'}{a(t')}
\end{equation}

If the light ray reaches $r=0$ at a time $t_0$, the Eq.(\ref{10}) gives
\begin{align}\label{11}
  0=&\sin^{-1}s(t_0)+\int_{t_L}^{t_0}\frac{\textrm dt}{a(t)}N(s(t)\hat{n},t)\nonumber\\
   =&\sin^{-1}r_L+\int_{t_L}^{t_0}\frac{\textrm dt}{a(t)}\left[N(s(t)\hat{n},t)-1\right]
\end{align}
A time interval $\delta t_L$ between the departure of successive light wave crests at the time $t_L$ of last scattering produces a time interval $\delta t_0$ between arrival of successive crests at $t_0$ given by the variation of Eq.(\ref{11}):
\begin{align}\label{12}
  0=&\frac{\delta t_L}{a(t_L)}\left[1-N(\rtl)+\int_{t_L}^{t_0}\frac{\sqrt{1-r^2}}{a(t)}\pdv{N(\rt)}{r}\dd t\right]\nonumber\\
  +&\frac{\delta t_L}{\sqrt{1-r_L^2}}\delta u_\gamma^r(\rtl)+\frac{\delta t_0}{a(t_0)}[N(0,t_0)-1]
\end{align}
where $\delta u_\gamma^r$ is the radial velocity of the photon gas or photon-electron-nucleon fluid arises from the change with time of radial coordinate of the light source. The total rate of change of quantity $N(\rtl)$ is
\begin{align}\label{13}
 \dv{t}N(\rt)&=\left(\pdv{t}N(\rt)\right)_{r=s(t)}\nonumber\\
 &-\frac{\sqrt{1-r^2}}{a}\left(\pdv{r}N(\rt)\right)_{r=s(t)}
\end{align}
so Eq.(\ref{12}) may be written as
\begin{align}\label{14}
  0=&\frac{\delta t_L}{a(t_L)}\left[1-N(0,t_0)+\int_{t_L}^{t_0}\pdv{N(\rt)}{t}\dd t\right]\nonumber\\
  +&\frac{\delta t_L}{\sqrt{1-r_L^2}}\delta u_\gamma^r(\rtl)+\frac{\delta t_0}{a(t_0)}[N(0,t_0)-1]
\end{align}
which to first order gives the ratio of coordinate time intervals
\begin{equation}\label{15}
  \fdv{t_L}{t_0}=\frac{a(t_L)}{a(t_0}\left[1-\int_{t_L}^{t_0}\pdv{N(\rt)}{t}\dd t-\frac{a(t_L)}{\sqrt{1-r_L^2}}\delta u_\gamma^r(\rtl)\right].
\end{equation}
But what we need is the ratio of the proper time intervals
\begin{equation}\label{16}
  \delta\tau_L=\sqrt{1+E(\rtl)}\,\delta t_L\qquad \delta\tau_0=\sqrt{1+E(0,t_0)}\,\delta t_0
\end{equation}
which to first order gives the ratio of the received and emitted frequencies as
\begin{align}\label{17}
  \frac{\upsilon_0}{\upsilon_L}=\frac{\delta\tau_L}{\delta\tau_0}=&\frac{a(t_L)}{a(t_0)}\bigg[1+\frac{1}{2}\left(E(\rtl)-E(0,t_0)\right)\nonumber\\
-& \int_{t_L}^{t_0}\pdv{N(\rt)}{t}\dd t-\frac{a(t_L)}{\sqrt{1-r_L^2}}\,\delta u_\gamma^r(\rtl)\bigg]
\end{align}
The temperature observed at the present time $t_0$ coming from direction $\hat{n}$ is
\begin{equation*}
  T(\hat{n})=\frac{\upsilon_0}{\upsilon_L}\left[\,\overline{T}(t_L)+\delta T(\rtl)\right]
\end{equation*}
In the absence of perturbations, the temperature observed in all directions would be
\begin{equation*}
  T_0=\frac{a(t_L)}{a(t_0)}\,\overline{T}(t_L)
\end{equation*}
So the fractional shift from its unperturbed value in the temperature observed coming from direction $\hat{n}$ is
\begin{align}\label{18}
    \frac{\Delta T(\hat{n})}{T_0}=&\frac{T(\hat{n})-T_0}{T_0}=\frac{1}{2}\left(E(\rtl)-E(0,t_0)\right)\nonumber\\
    -&\int_{t_L}^{t_0}\pdv{N(\rt)}{t}\dd t-\frac{a(t_L)}{\sqrt{1-r_L^2}}\delta u_\gamma^r(\rtl)\nonumber\\
    +&\,\frac{\delta T(\rtl)}{\overline{T}(t_L)}
\end{align}\label{18}
For scalar perturbation $ h_{rr}$ we have
\begin{equation*}
  h_{rr}=a^2\left(\frac{A}{1-r^2}+H_{rr}\,B\right)
\end{equation*}
also, the radial fluid velocity can be expressed in terms of velocity potential $\delta u_\gamma$ as
\begin{equation*}
  \delta u_\gamma^r=g^{r\mu}\nabla_\mu\delta u_\gamma=\frac{1-r^2}{a^2}\pdv{r}\delta u_\gamma
\end{equation*}
thus the scalar contribution to the temperature fluctuation is
\begin{align}\label{19}
  \left(\frac{\Delta T(\hat{n})}{T_0}\right)^S=&\frac{1}{2}\left(E(\rtl)-E(0,t_0)\right)-\int_{t_L}^{t_0}\pdv{N(\rt)}{t}\dd t\nonumber\\
  -&\frac{\sqrt{1-r_L^2}}{a(t_L)}\left(\pdv{\delta u_\gamma(\rt)}{r}\right)_{r=r_L}+\frac{\delta T(\rtl)}{\overline{T}(t_L)}
\end{align}
where
\begin{equation*}
  N=\frac{1}{2}\left(A+(1-r^2)H_{rr}\,B-E\right).
\end{equation*}

It will be much more convenient to rewrite Eq.(\ref{19}) so that each term in the integral being gauge invariant. For this purpose, we may use the identity
\begin{align*}
  (1-r^2)\pdv{t}(H_{rr}\,B)=&-\dv{t}\left[a^2\ddot{B}+a\dot{a}\dot{B}+\sqrt{1-r^2}\,a\pdv{\dot{B}}{r}\right]\nonumber\\
  +&\pdv{t}(a^2\ddot{B}+a\dot{a}\dot{B})
\end{align*}
This gives for the integrand in Eq.(\ref{19})
\begin{align*}
  \pdv{N(\rt)}{t}=&-\frac{1}{2}\dv{t}\left[a^2\ddot{B}+a\dot{a}\dot{B}+\sqrt{1-r^2}\,a\pdv{\dot{B}}{r}\right]\nonumber\\
  +&\frac{1}{2}\pdv{t}(a^2\ddot{B}+a\dot{a}\dot{B}+A-E)
\end{align*}
Therefore the scalar temperature fluctuations may be written as
\begin{align}\label{20}
  \left(\frac{\Delta T(\hat{n})}{T_0}\right)^S=& \left(\frac{\Delta T(\hat{n})}{T_0}\right)^S_{early}+ \left(\frac{\Delta T(\hat{n})}{T_0}\right)^S_{late}\nonumber\\
  +& \left(\frac{\Delta T(\hat{n})}{T_0}\right)^S_{ISW}
\end{align}
where
\begin{align}\label{21}
  \left(\frac{\Delta T(\hat{n})}{T_0}\right)^S_{early}=&-\frac{1}{2}a^2(t_L)\ddot{B}(\rtl)\nonumber\\
  &-\frac{1}{2}a(t_L)\dot{a}(t_L)\dot{B}(\rtl)\nonumber\\
  &+\frac{1}{2}E(\rtl)+\frac{\delta T(\rtl)}{\overline{T}(t_L)}\nonumber\\
  &-\sqrt{1-r_L^2}\,a(t_L)\times\nonumber\\
  &\left[\pdv{r}(\frac{1}{2}\dot{B}(\rt)+\frac{\delta u_\gamma(\rt)}{a^2(t_L)})\right]_{r=r_L}
\end{align}

\begin{align}\label{22}
  \left(\frac{\Delta T(\hat{n})}{T_0}\right)^S_{late}=&\frac{1}{2}a^2(t_0)\ddot{B}(0,t_0)+\frac{1}{2}a(t_0)\dot{a}(t_0)\dot{B}(0,t_0)\nonumber\\
  +&\frac{1}{2}a(t_0)\left[\pdv{r}\dot{B}(\rt)\right]_{r=0}-\frac{1}{2}E(0,t_0)
\end{align}

\begin{align}\label{23}
  \left(\frac{\Delta T(\hat{n})}{T_0}\right)^S_{ISW}&=-\frac{1}{2}\int_{t_L}^{t_0}\dd t\Bigg[\pdv{t}\Big(a^2(t)\ddot{B}(\rt)\nonumber\\
  +&a(t)\dot{a}(t)\dot{B}(\rt)+A(\rt)-E(\rt)\Big)\Bigg]_{r=s(t)}
\end{align}
so, in synchronous gauge (where $E=~0$), temperature fluctuations may be expanded as
\begin{align}\label{24}
  \left(\frac{\Delta T(\hat{n})}{T_0}\right)^S_{early}=\sum_{n\ell m}\alpha_{\ell m}\bigg[&F_n\Pi_{n\ell}(\chi_L)\nonumber\\
  +&G_n\,\dv{\chi_L}\Pi_{n\ell}(\chi_L)\bigg]Y_{\ell m}(\theta,\varphi)
\end{align}
\begin{align}\label{25}
   &\left(\frac{\Delta T(\hat{n})}{T_0}\right)^S_{ISW}=-\frac{1}{2}\int_{t_L}^{t_0}\dd t\sum_{n\ell m}\alpha_{\ell m}\dv{t}\bigg(a^2(t)\ddot{B}_n(t)\nonumber\\
   &+a(t)\dot{a}(t)\dot{B}_n(t)+A_n(t)-E_n(t)\bigg)\,\mathcal{Y}_{n\ell m}(\chi(t),\theta,\varphi)
\end{align}
where
\begin{equation*}
  F_n=-\frac{1}{2}a^2(t_L)\ddot{B}(t_L)-\frac{1}{2}a(t_L)\dot{a}(t_L)\dot{B}_n(t_L)+\frac{\delta T_n(t_L)}{\overline{T}(t_L)}
\end{equation*}
\begin{equation*}
  G_n=-\left(\frac{1}{2}a(t_L)\dot{B}(t_L)+\frac{\delta u_n(t_L)}{a(t_L)}\right)
\end{equation*}

\section{Temperature fluctuations using the Boltzmann equation formalism}\label{504}
In this section we extract the temperature fluctuation again but this time from the Boltzmann equation.\\
The Boltzmann equation that governs the evolution of the distribution of photons in phase space, can be written as
\begin{align*}
  \pdv{n_\gamma^{ij}}{t}&+\pdv{n_\gamma^{ij}}{x^k}\frac{p^k}{p^0}+\pdv{n_\gamma^{ij}}{p_k}\frac{p^lp^m}{2p^0}\pdv{g_{lm}}{x^k}\nonumber\\
   &+\left(\Gamma^i_{k\lambda}-\frac{p_i}{p^0}\Gamma^0_{k\lambda}\right)\frac{p^\lambda}{p^0}n^{kj}_\gamma\nonumber\\
   &+\left(\Gamma^j_{k\lambda}-\frac{p_j}{p^0}\Gamma^0_{k\lambda}\right)\frac{p^\lambda}{p^0}n^{ki}_\gamma=C^{ij}
\end{align*}
where $n_\gamma^{ij}$ is the number density matrix for photons and $C^{ij}$ is a term representing the effect of photon scattering.
Let us introduce the dimensionless intensity matrix as follows \cite{r20}
\begin{equation*}
  a^4(t)\overline{\rho}_\gamma(t)J^{ij}(\textbf{x},\hat{p},t)\equiv a^2(t)\int_{0}^{\infty}\delta n_\gamma^{ij}(\textbf{x},p\hat{p},t)4\pi\,p^3 \dd p
\end{equation*}
One can gets the Boltzmann equation in terms of $J^{ij}(\textbf{x},\hat{p},t)$ matrix as
\begin{align}\label{26}
 &\pdv{t}J^{ij}(\textbf{x},\hat{p},t)
 +\frac{\hat{p}^k}{a(t)}\pdv{x^k}J^{ij}(\textbf{x},\hat{p},t)\nonumber\\
 +&\frac{\hat{p}^s}{a(t)}\left(\widetilde{\Gamma}^i_{ks}J^{kj}(\textbf{x},\hat{p},t)+\widetilde{\Gamma}^j_{ks}J^{ki}(\textbf{x},\hat{p},t)\right)\nonumber\\
 +&\hat{p}^s\hat{p}^t\pdv{t}\left(a^{-2}\delta g_{st}\right)\left(\tilde{g}^{ij}-\hat{p}^i\hat{p}^j\right)\nonumber\\
 -&2\,\frac{\hat{p}^k\hat{p}^s\hat{p}^t}{a(t)}\partial_k\,\tilde{g}_{st}J^{ij}(\textbf{x},\hat{p},t)\nonumber\\
 =&-\omega_c(t)J^{ij}(\textbf{x},\hat{p},t)+\frac{2\,\omega_c(t)}{a(t)}\tilde{g}^{kl}\hat{p}_l\,\delta u_k(\textbf{x},t)\left(\tilde{g}^{ij}-\hat{p}^i\hat{p}^j\right)\nonumber\\
 +&\frac{3\,\omega_c(t)}{8\pi}\int \dd[2]{\hat{p}_1}\sqrt{Det\,\tilde{g}^{ij}}\bigg[J^{ij}(\textbf{x},\hat{p}_1,t)\nonumber\\
 -&\tilde{g}^{ik}\hat{p}_k\hat{p}_lJ^{jl}(\textbf{x},\hat{p}_1,t)-\tilde{g}^{jk}\hat{p}_k\hat{p}_lJ^{il}(\textbf{x},\hat{p}_1,t)\nonumber\\
 +&\tilde{g}^{ik}\tilde{g}^{jl}\hat{p}_k\hat{p}_l\hat{p}_m\hat{p}_nJ^{mn}(\textbf{x},\hat{p}_1,t)\bigg]
\end{align}
where $\omega_c(t)$ is the collision rate of a photon with electrons in the baryonic plasma and $\delta u_l(\textbf{x},t)$ is the peculiar velocity of the baryonic plasma. The term containing $\delta u_l (\textbf{x},t)$ will be added to the Boltzmann equation in scalar or vector modes.\\
Now using the perturbation theory in a spatially closed universe, we expand the metric perturbations, $J^{ij} (\textbf{x},\hat{p},t)$ and $\delta u_l (\textbf{x},t)$ in terms of the eigenvalues of Laplace-Beltrami operator. The metric perturbation in scalar mode can be written as
\begin{equation*}
  a^{-2}(t)\delta g_{st}=A\,\tilde{g}_{st}+H_{st}\,B
\end{equation*}
where for the perturbative quantities $A$ and $B$ we can write
\begin{align*}
 A(\textbf{x},t)&=\sum_{n\ell m}\alpha_{\ell m}A_n(t)\mathcal{Y
 }_{n\ell m}(\chi,\theta,\phi)\\
 B(\textbf{x},t)&=\sum_{n\ell m}\alpha_{\ell m}B_n(t)\mathcal{Y
 }_{n\ell m}(\chi,\theta,\phi)
\end{align*}
The plasma velocity can be expressed in terms of the velocity potential and then expand as
\begin{equation*}
 \delta u_k(\textbf{x},t)=\nabla_k\,\delta u(\textbf{x},t)=\sum_{n\ell m}\alpha_{\ell m}\delta u_n(t)\nabla_k\,\mathcal{Y}_{n\ell m}(\chi,\theta,\phi)
\end{equation*}
We may write the $J^{ij} (\textbf{x},\hat{p},t)$  matrix as
\begin{align}\label{27}
  J^{ij} (\textbf{x},\hat{p},t)=&\sum_{n\ell m}\alpha_{\ell m}\bigg[\frac{1}{2}\left(\Delta_{Tn}(t)-\Delta_{Pn}(t)\right)\times\nonumber\\
  &\left(\tilde{g}^{ij}-\hat{p}^i\hat{p}^j\right)\mathcal{Y}_{n\ell m}(\chi,\theta,\phi)\nonumber\\
  +&\,\Delta_{Pn}(t)k_n^{-2}q^{ij}\mathcal{Y}_{n\ell m}(\chi,\theta,\phi)\bigg]
\end{align}
where
\begin{equation*}
  q^{ij}=\frac{(\nabla^i-\hat{p}^i\hat{p}_s\nabla^s)(\nabla^j-\hat{p}^j\hat{p}_t\nabla^t)}{\mathcal{Y}^{-1}_{n\ell m}(\chi,\theta,\phi)k_n^{-2}(\nabla^2-\hat{p}_s\hat{p}_tH^{st})\mathcal{Y}_{n\ell m}(\chi,\theta,\phi)}
\end{equation*}
Note that the trace $J^i_{\;i}(\textbf{x},\hat{p},t)$, which we will derive the temperature fluctuation from it, equals to
\begin{equation*}
  J^i_{\;i}(\textbf{x},\hat{p},t)=\sum_{n\ell m}\alpha_{\ell m}\Delta_{Tn}(t)\mathcal{Y}_{n\ell m}(\chi,\theta,\phi)
\end{equation*}
Also, we introduce the source functions $\varphi_n (t)$ and $\mathcal{J}_n (t)$ as
\begin{align*}
 \int\frac{\dd[2]{\hat{p}_1}}{4\pi}&\sqrt{Det\,\tilde{g}^{ij}}J^{ij}(x,p_1,t)\\
 &=\sum_{n\ell m}\alpha_{\ell m}\Big[\varphi_n(t)\tilde{g}^{ij}\mathcal{Y}_{n\ell m}(\chi,\theta,\phi)\\
 &-\frac{1}{2}\mathcal{J}_n(t)H^{ij}k_n^{-2}\mathcal{Y}_{n\ell m}(\chi,\theta,\phi)\Big]
\end{align*}

In this particular coordinate $(\chi,\theta,\phi)$, the momentum $\hat{p}$ for the photon coming from direction $\hat{n}$ will be $ \hat{p}=-\hat{n}=(-1,0 ,0)=-e_\chi$.\\
Because of the conditions $\hat{p}_i\,J^{ij}(\textbf{x},\hat{p},t)=0$ and\\ $\hat{p}_j\,J^{ij}(\textbf{x},\hat{p},t)=0$, this is just the $\chi\chi$ element of the $J^{ij}(\textbf{x},\hat{p},t)$ matrix that contributes to the fluctuation calculations. So, for $\hat{p}^s\hat{p}^tH_{st},\; \hat{p}^k\nabla_k$ and\\ $\hat{p}^k\hat{p}^m\hat{p}^n\partial_k\tilde{g}_{mn}$ we can write
\begin{align*}
 &\hat{p}^k\nabla_k=-\nabla_\chi\qquad\hat{p}^s\hat{p}^tH_{st}=\nabla_\chi\nabla_\chi\\
 &\hat{p}^k\hat{p}^m\hat{p}^n\partial_k\tilde{g}_{mn}=-\partial_\chi \tilde{g}_{\chi\chi}=0
\end{align*}
Inserting all above relations in Eq.(\ref{26}), one can show that the Boltzmann equation for the matrix\\
$J^{ij}(\textbf{x},\hat{p},t)$ yields two coupled Boltzmann equations for $\Delta_{Tn}(t)$ and $\Delta_{Pn}(t)$ as
\begin{align}\label{28}
  &\dot{\Delta}_{Tn}(t)\Pi_{n \ell}(\chi)-\Delta_{Tn}(t)\frac{1}{a(t)}\dv{\chi}\Pi_{n \ell}(\chi)\nonumber\\
  &+2\dot{A}_n(t)\Pi_{n \ell}(\chi)+2\dot{B}_n(t)\dv[2]{\chi}\Pi_{n \ell}(\chi)\nonumber\\
  &=-\omega_c(t)\Delta_{Tn}(t)\Pi_{n\ell}(\chi)+3\,\omega_c(t)\varphi_n(t)\Pi_{n\ell}(\chi)\nonumber\\
  &-\frac{4\,\omega_c(t)}{a(t)}\delta u_n(t)\dv{\chi}\Pi_{n\ell}(\chi)\nonumber\\
  &+\frac{3}{4}\omega_c(t)\mathcal{J}_n(t)\left(1+k_n^{-2}\dv[2]{\chi}\right)\Pi_{n\ell}(\chi)
\end{align}

\begin{align}\label{29}
  &\dot{\Delta}_{Pn}(t)\Pi_{n \ell}(\chi)-\Delta_{Pn}(t)\frac{1}{a(t)}\dv{\chi}\Pi_{n \ell}(\chi)\nonumber\\
  &=-\omega_c(t)\Delta_{Pn}(t)\Pi_{n\ell}(\chi)\nonumber\\
  &+\frac{3}{4}\omega_c(t)\mathcal{J}_n(t)\left(1+k_n^{-2}\dv[2]{\chi}\right)\Pi_{n\ell}(\chi)
\end{align}
where again $\Pi_{n\ell} (\chi)$ is the hyperspherical Bessel function that has been introduced in section \ref{502}. Steps for this derivation could be found in Appendix.\\
For a photon coming from direction $\hat{n}$ we have $\dd t=-a(t)\dd \chi$, so
\begin{align*}
  \dv{\chi}&=-a(t)\dv{t}\nonumber\\
  \dv[2]{\chi}&=-a(t)\dot{a}(t)\dv{t}+a^2(t)\dv[2]{t}\nonumber\\
  \chi(t)&=\int_{t}^{t_0}\frac{\dd t'}{a(t')}
\end{align*}
By using above relations and also integrating by parts of Eqs.(\ref{28}),(\ref{29}), one can get the line of sight solution for $\Delta_{Tn}(t)$ as
\begin{subequations}\label{30}
  \begin{equation}\label{30-a}
  \Delta_{Tn}(t)\Pi_{n\ell} (\chi)=\int_{t_1}^{t}\exp\left(-\int_{t'}^{t}\dd t''\omega_c(t'')\right)\mathcal{G}_{n \ell}(t')\dd t'
\end{equation}
\begin{align}\label{30-b}
 \mathcal{G}_{n\ell}=&\left(-2\dot{A}_n+3\,\omega_c(t)\varphi_n(t)+\frac{3}{4}\omega_c(t)\mathcal{J}_n(t)\right)\Pi_{n\ell}(t)\nonumber\\
 +&\bigg(4\,\omega_c(t)\delta u_n(t)-2a\dot{a}\dot{B}_n\nonumber\\
 +&\frac{3}{4}\omega_c(t)\,\mathcal{J}_n(t)\,a\dot{a}k_n^{-2}\bigg)\dv{t}\Pi_{n\ell}(t)\nonumber\\
 +&\left(-2a^2\dot{B}_n+\frac{3}{4}\omega_c(t)\mathcal{J}_n(t)a^2k_n^{-2}\right)\dv[2]{t}\Pi_{n\ell}(t)
\end{align}
\end{subequations}
where $t_1$ is a time that we choose it to be sufficiently early, so that $\omega_c (t_1 )$ is much bigger than the expansion rate of the universe and $t$ at any time after recombination.
Also the line of sight solution for $\Delta_{Pn}(t)$ would be
\begin{subequations}\label{31}
  \begin{equation}\label{31-a}
  \Delta_{Pn}(t)\Pi_{n\ell} (\chi)=\int_{t_1}^{t}\exp\left(-\int_{t'}^{t}\dd t''\omega_c(t'')\right)\mathcal{H}_{n \ell}(t')\dd t'
\end{equation}
\begin{align}\label{31-b}
 \mathcal{H}_{n\ell}=&\frac{3}{4}\omega_c(t)\mathcal{J}_n(t)\bigg(\Pi_{n\ell}(t)+k_n^{-2}a\dot{a}\dv{t}\Pi_{n\ell}(t)\nonumber\\
 +&k_n^{-2}a^2\dv[2]{t}\Pi_{n\ell}(t)\bigg)
\end{align}
\end{subequations}
The temperature fluctuation at our position $\textbf{x}=0$ and time $t=t_0$ can be written as
\begin{align}\label{34}
\frac{\Delta T(\hat{n})}{T_0}=&\frac{1}{4}J^i_{\;i}(\textbf{x}=0,\hat{p},t_0)\nonumber\\
=&\frac{1}{4}\sum_{n\ell m}\alpha_{\ell m}\Delta_{Tn}(t_0)\mathcal{Y}_{n\ell m}(\chi=0,\theta,\phi)\nonumber\\
=&\frac{1}{4}\sum_{n\ell m}\alpha_{\ell m}\Delta_{Tn}(t_0)\Pi(t_0)Y_{\ell m}(\theta,\phi)\nonumber\\
=&\sum_{\ell m}a^S_{T,\ell m}Y_{\ell m}(\theta,\phi)
\end{align}
where the scalar contribution to the multipole coefficient  $a_{T,\ell m}^S$ is
\begin{equation}\label{35}
  a^S_{T,\ell m}=\frac{1}{4}\sum_{n}\alpha_{\ell m}\int_{t_1}^{t_0}\exp\left(-\int_{t}^{t_0}\dd t'\omega_c(t')\right)\mathcal{G}_{n\ell}(t)\dd t
\end{equation}

By assuming that $\omega_c (t)$ drops sharply at time $t_L$ from a value much greater than the expansion rate to zero (a sudden transition from opacity to transparency), the integral $\int_{t_1}^{t_0}\dd t\,\omega_c(t)\exp\left(-\int_{t}^{t_0}\dd t'\,\omega_c(t')\right)$ is non zero only in a narrow interval around  $t_L$. Also under the same assumption, the factor $\exp\left(-\int_{t}^{t_0}\dd t'\,\omega_c(t')\right)$ rise sharply from zero for time before recombination to unity for $t>t_L$.
Using above approximation and integrating by parts, one can show that $a_{T,\ell m}^S$ can be written as
\begin{subequations}\label{36}
  \begin{equation}\label{36-a}
  a^S_{T,\ell m}=\left(a^S_{T,\ell m}\right)_{early}+\left(a^S_{T,\ell m}\right)_{ISW}
\end{equation}
\begin{align}\label{36-b}
\left(a^S_{T,\ell m}\right)_{early}=&\sum_{n}\alpha_{\ell m}\Bigg[\bigg(\frac{3}{4}\varphi_n(t_L)-\frac{1}{2}a^2(t_L)\ddot{B}_n(t_L)\nonumber\\
-&\frac{1}{2}a(t_L)\dot{a}(t_L)\dot{B}_n(t_L)+\frac{3}{16}\mathcal{J}_n(t_L)\bigg)\Pi_{n\ell}(t_L)\nonumber\\
+&\bigg(\delta u_n(t_L)+\frac{1}{2}a^2(t_L)\dot{B}_n(t_L)\nonumber\\
+&\frac{3}{16}\mathcal{J}_n(t_L)k_n^{-2}a(t_L)\dot{a}(t_L)\bigg)\dv{t_L}\Pi_{n\ell}(t_L)\nonumber\\
+&\frac{3}{16}\mathcal{J}_n(t_L)k_n^{-2}a^2(t_L)\dv[2]{t_L}\Pi_{n\ell}(t_L)\Bigg]
\end{align}
\begin{align}\label{36-c}
\left(a^S_{T,\ell m}\right)_{ISW}&=-\frac{1}{2}\sum_{n}\alpha_{\ell m}\int_{t_L}^{t_0}\Pi_{n\ell}(t)\Bigg[\dv{t}\bigg(A_n(t)\nonumber\\
&+a^2(t)\ddot{B}_n(t)+a(t)\dot{a}(t)\dot{B}_n(t)\bigg)\Bigg]\dd t
\end{align}
\end{subequations}
In local thermal equilibrium, photons are unpolarized and have an isotropic momentum distribution. Therefore similar to the flat case \cite{r20} we will have
\begin{equation*}
  \mathcal{J}_n(t_L)=0\qquad\varphi_n(t_L)=\frac{4}{3}\frac{\delta T_n(t_L)}{\overline{T}}
\end{equation*}
Using above relations and replacing the time derivative with a $\chi$ derivative using the relation $\dv{\chi_L}=-a(t_L)\dv{t_L}$, one can derive the final formulas for the scalar mode temperature fluctuation as
\begin{align}\label{39}
  &\left(\frac{\Delta T(\hat{n})}{T_0}\right)^S_{early}=\nonumber\\
  &\sum_{n\ell m}\alpha_{\ell m}\bigg[F_n\Pi_{n\ell}(\chi_L)+G_n\dv{\chi_L}\Pi_{n\ell}\bigg]Y_{\ell m}(\theta,\phi)
\end{align}
\begin{align}\label{40}
  \left(\frac{\Delta T(\hat{n})}{T_0}\right)^S_{ISW}&=-\frac{1}{2}\sum_{n\ell m}\alpha_{\ell m}\int_{t_L}^{t_0}\dd t\,\Pi_{n\ell}(t)\Bigg[\dv{t}\bigg(A_n(t)\nonumber\\
  &+a^2(t)\ddot{B}_n(t)+a(t)\dot{a}(t)\dot{B}_n(t)\bigg)\Bigg]Y_{\ell m}(\theta,\phi)
\end{align}
where
\begin{equation*}
  F_n=-\frac{1}{2}a^2(t_L)\ddot{B}(t_L)-\frac{1}{2}a(t_L)\dot{a}(t_L)\dot{B}_n(t_L)+\frac{\delta T_n(t_L)}{\overline{T}(t_L)}
\end{equation*}
\begin{equation*}
  G_n=-\left(\frac{1}{2}a(t_L)\dot{B}(t_L)+\frac{\delta u_n(t_L)}{a(t_L)}\right)
\end{equation*}
This was also derived at the end of section \ref{503} by following photon trajectories after the time of last scattering.

At the next section, in order to extract a simple analytic formula for the multipole coefficient, we neglect the integrated Sachs-Wolfe effect that is important only for relatively small values of $\ell$, where cosmic variance set a limit on the accuracy with which we can measure the multipole coefficient.

\section{Temperature multipole coefficient at the hydrodynamic limit}\label{505}
Now, we calculate the multipole coefficient of angular temperature correlations using the general formula for the temperature fluctuation derived at the end of the previous section.
We can expand temperature fluctuation at some instant of time in terms of orthogonal functions $\mathcal{Y}_{n\ell m} (\chi,\theta,\phi)$ as
\begin{align}\label{41}
  \Delta T(\hat{n})=&\sum_{n\ell m}\Delta T_{n\ell m}\,\mathcal{Y}_{n\ell m}(\chi,\theta,\phi)\nonumber\\
  =&\sum_{n\ell m}\alpha_{\ell m}\Delta T_{n}\,\mathcal{Y}_{n\ell m}(\chi,\theta,\phi)
\end{align}
where $\Delta T_{n\ell m}$ just like $\Delta T(\hat{n})$ is a scalar random field defined on $S^3 (\alpha)$. The mean value product of $\Delta T(\hat{n})$ with itself called the two-point covariance function and can be written as
\begin{align}\label{42}
  &\langle\Delta T(\hat{n})\Delta T^\ast(\hat{n}')\rangle=\nonumber\\
  &\sum_{n\ell m}\sum_{n'\ell' m'}\langle\Delta T_{n\ell m}\Delta T^\ast_{n'\ell' m'}\rangle\mathcal{Y}_{n\ell m}(\chi,\hat{n})\mathcal{Y}^\ast_{n'\ell' m'}(\chi',\hat{n}')
\end{align}
On the other hand, homogeneity and isotropy imply that
\begin{subequations}\label{43}
  \begin{align}\label{43-a}
 \langle\Delta T_{n\ell m}\Delta T^\ast_{n'\ell' m'}\rangle
 =&\langle\alpha_{\ell m}\alpha_{\ell'  m'}\rangle\langle\Delta T_n \Delta T^\ast_{n'}\rangle\nonumber\\
 =&\mathcal{P}_{\Delta T}(n)\delta_{nn'}\delta_{\ell \ell'}\delta_{mm'}\qquad
\end{align}
\begin{align}\label{43-b}
 \mathcal{P}_{\Delta T}(n)=|\Delta T_n|^2
\end{align}
\end{subequations}
so we can write the Eq.(\ref{42}) as
\begin{align}\label{45}
  &\langle\Delta T(\hat{n})\Delta T^\ast(\hat{n}')\rangle=\nonumber\\
  &\sum_{n\ell m}\,\mathcal{P}_{\Delta T}(n)\Pi_{n\ell}(\chi)\Pi_{n\ell}(\chi')Y_{\ell m}(\hat{n})Y_{\ell m}(\hat{n}')
\end{align}
By using the relation $\sum_{m}Y_{\ell m}(\hat{n})Y_{\ell m}(\hat{n}')=\\\frac{2\ell+1}{4\pi}P_{\ell}(\hat{n}.\hat{n}')$ we have
\begin{align}\label{46}
  &\langle\Delta T(\hat{n})\Delta T^\ast(\hat{n}')\rangle=\nonumber\\
  &\frac{1}{4\pi}\sum_{n\ell}(2\ell+1)\mathcal{P}_{\Delta T}(n)\Pi_{n\ell}(\chi)\Pi_{n\ell}(\chi')P_\ell(\hat{n}.\hat{n}').
\end{align}

For the multipole coefficient of angular temperature correlations we have
\begin{align}\label{47}
  C^S_{TT,\ell}=\frac{1}{4\pi}\int\dd[2]{\hat{n}}\int\dd[2]{\hat{
  n}'}P_\ell(\hat{n}.\hat{n}')\langle\Delta T(\hat{n})\Delta T^\ast(\hat{n}')\rangle
\end{align}
By inserting Eq.(\ref{46}) into the above equation and using $\int\dd[2]{\hat{n}'}P_\ell(\hat{n}.\hat{n}')\,P_{\ell'}(\hat{n}.\hat{n}')=\frac{4\pi}{2\ell+1}\delta_{\ell\ell'}$ and $ \chi=\chi'=\chi_L$, one can show that
\begin{align}\label{48}
  C^S_{TT,\ell}=\sum_{n}\mathcal{P}_{\Delta T}(n)\Pi^2_{n\ell}(\chi_L)=\sum_{n}|\Delta T_n|^2\Pi^2_{n\ell}(\chi_L)
\end{align}
By comparing Eq.(\ref{41}) with Eq.(\ref{39}), the temperature multipole coefficient can be written as
\begin{subequations}\label{49}
  \begin{align}\label{49a}
  C^S_{TT,\ell}=T_0^2\sum_{n}\bigg[F_n\Pi_{n\ell}(\chi_L)+G_n\dv{\chi_L}\Pi_{n\ell}(\chi_L)\bigg]^2
\end{align}
where, as before
\begin{equation}\label{49b}
  F_n=-\frac{1}{2}a^2(t_L)\ddot{B}(t_L)-\frac{1}{2}a(t_L)\dot{a}(t_L)\dot{B}_n(t_L)+\frac{\delta T_n(t_L)}{\overline{T}(t_L)}
\end{equation}
\begin{equation}\label{49c}
  G_n=-\left(\frac{1}{2}a(t_L)\dot{B}(t_L)+\frac{\delta u_n(t_L)}{a(t_L)}\right)
\end{equation}
\end{subequations}

The hyperspherical Bessel function $\Pi_{n\ell} (\chi)$ satisfy
\begin{align}\label{50}
  \dv[2]{\Pi_{n\ell}(\chi)}{\chi}+&2\cot\chi\dv{\Pi_{n\ell}(\chi)}{\chi}\nonumber\\
  +&\left[(n^2-1)-\frac{\ell(\ell+1)}{\sin^2\chi}\right]{\Pi_{n\ell}(\chi)}=0
\end{align}
By introducing $U_{n\ell} (\chi)=\Pi_{n\ell} (\chi)  \sin\chi$, above equation can be written in a more useful form as
\begin{align}\label{51}
  \dv[2]{U_{n\ell}(\chi)}{\chi}=\left[\frac{\ell(\ell+1)}{\sin^2\chi}-n^2\right]{U_{n\ell}(\chi)}
\end{align}
that is suitable for employing the WKB approximation as follows.\\
Dividing both sides of above equation with $\ell(\ell+1)$, we have at large $\ell$
\begin{align}\label{52}
  \frac{1}{\ell^2}\dv[2]{U_{n\ell}(\chi)}{\chi}=\left[\frac{1}{\sin^2\chi}-\frac{n^2}{\ell^2}\right]U_{n\ell}(\chi)=Q(\chi)\,U_{n\ell}(\chi)
\end{align}
Its solution has exponential behavior in regions where $Q>0$ and oscillatory behavior when $Q<0$. So for relatively large $\ell$ and $n>\frac{\ell}{\sin\chi}$, we have \cite{r21}
\begin{align}\label{53}
  U_{n\ell}(\chi)\cong C[-Q]^{-1/4}\sin\left[\ell\int_{\chi_0}^{\chi}\sqrt{-Q(t)}\dd t+\frac{\pi}{4}\right]
\end{align}
where $C$ is a normalization constant and will be determined so as to match the normalization of ordinary spherical bessel function in the flat space limit.\\
So the WKB approximation of hyperspherical Bessel function $\Pi_{n\ell}(\chi)$ will be
\begin{align}\label{54}
  \Pi_{n\ell} (\chi)=4\pi\sqrt{\frac{n}{\ell\alpha^3}}\,\frac{\sin\left[\ell\int_{\chi_0}^{\chi}\sqrt{-Q(t)}\dd t+\frac{\pi}{4}\right]}{\sin\chi\left(\frac{n^2}{\ell^2}-\frac{1}{\sin^2\chi}\right)^{\frac{1}{4}}}
\end{align}
For large $\ell$ the phase $\ell\int_{\chi_0}^{\chi}\sqrt{-Q(t)}\dd t$ in above equation is a very rapidly increasing function of $\chi$ , so the derivative acting on the hyperspherical Bessel function in Eq.(\ref{49}) can be taken to act chiefly on this phase
\begin{align}\label{55}
  &\dv{\Pi_{n\ell} (\chi)}{\chi}=\nonumber\\
  &4\pi\sqrt{\frac{n}{\ell\alpha^3}}\,\frac{\ell\left(\frac{n^2}{\ell^2}-\frac{1}{\sin^2\chi}\right)^{\frac{1}{2}}\cos\left[\ell\int_{\chi_0}^{\chi}\sqrt{-Q(t)}\dd t+\frac{\pi}{4}\right]}{\sin\chi\left(\frac{n^2}{\ell^2}-\frac{1}{\sin^2\chi}\right)^{\frac{1}{4}}}
\end{align}
By using above WKB aproximation, the multipole coefficient of temperature fluctuation takes the form
\begin{align}\label{56}
   &C^S_{TT,\ell}=\nonumber\\
   &\frac{16\pi^2\,T_0^2}{\alpha^3}\sum_{n>\frac{\ell}{\sin\chi_L}}\frac{n}{\ell\, (\frac{n^2}{\ell^2}-\frac{1}{\sin^2\chi_L})^{\frac{1}{2}}\sin^2\chi_L}\times\nonumber\\
   &\Bigg[F_n\,\sin\left[\ell\int_{\chi_0}^{\chi}\sqrt{-Q(t)}\dd t+\frac{\pi}{4}\right]\nonumber\\
   &+\ell\,(\frac{n^2}{\ell^2}-\frac{1}{\sin^2\chi_L})^{\frac{1}{2}}\,G_n\cos\left[\ell\int_{\chi_0}^{\chi}\sqrt{-Q(t)}\dd t+\frac{\pi}{4}\right]\Bigg]^2
\end{align}
For large $\ell$, the functions $\sin\left[\ell\int_{\chi_0}^{\chi}\sqrt{-Q(t)}\dd t+\frac{\pi}{4}\right]$ and $\cos\left[\ell\int_{\chi_0}^{\chi}\sqrt{-Q(t)}\dd t+\frac{\pi}{4}\right]$ oscillate very rapidly, so the functions $\sin^2\left[\ell\int_{\chi_0}^{\chi}\sqrt{-Q(t)}\dd t+\frac{\pi}{4}\right]$ and\\ $\cos^2\left[\ell\int_{\chi_0}^{\chi}\sqrt{-Q(t)}\dd t+\frac{\pi}{4}\right]$ average to $\frac{1}{2}$, while the function $\sin\left[\ell\int_{\chi_0}^{\chi}\sqrt{-Q(t)}\dd t+\frac{\pi}{4}\right]\cos\left[\ell\int_{\chi_0}^{\chi}\sqrt{-Q(t)}\dd t+\frac{\pi}{4}\right]$ averages to zero. So the Eq.(\ref{56}) becomes
\begin{align}\label{57}
   &C^S_{TT,\ell}=\nonumber\\
   &\frac{8\pi^2\,T_0^2}{\alpha^3}\sum_{n>\frac{\ell}{\sin\chi_L}}\frac{n}{ \ell\,(\frac{n^2}{\ell^2}-\frac{1}{\sin^2\chi_L})^{\frac{1}{2}}\sin^2\chi_L}\times\nonumber\\
   &\Bigg[F_n^2+\ell^2(\frac{n^2}{\ell^2}-\frac{1}{\sin^2\chi_L})\,G_n^2\Bigg]
\end{align}

In order to find $F_n$ and $G_n$ in terms of known perturbations we notice that, until near the time of recombination, the rate of collisions of photons with free electrons was so great that photons were in local thermal equilibrium with the baryonic plasma, and so can be treated hydrodynamically. To be specific, in this section and forthcoming calculations we neglect anisotropy inertia.\\

In synchronous gauge, the energy-momentum and field equations govern the perturbations are \cite{r15}
\begin{align}\label{58}
  \dv{t}(a^2\psi)=-4\pi Ga^2(\delta \rho+3\,\delta P+\nabla^2\Pi^S)
\end{align}
\begin{align}\label{59}
  \delta \dot{\rho}+&3\frac{\dot{a}}{a}(\delta \rho+\delta P)+\frac{1}{a^2}\nabla^2\left((\bar{\rho}+\bar{P})\delta u+a\dot{a}\Pi^S\right)\nonumber\\
  +&(\bar{\rho}+\bar{P})\psi=0
\end{align}
\begin{align}\label{60}
  \delta P+&\partial_0\left((\bar{\rho}+\bar{P})\delta u\right)+3\frac{\dot{a}}{a}(\bar{\rho}+\bar{P})\delta u+\nabla^2\Pi^S\nonumber\\
  +&2K\Pi^S=0
\end{align}
By considering $\Pi^S=0$, the equations format in non-flat universe is the same as that ones in flat universe, so we can use the results derived in \cite{r20} for the perturbations just by changing $q^2$ in to $k_n^2=q_n^2-1/\alpha^2$ for the spatially closed universe case. It is noted that for a closed universe with a very small amount of curvature, $k_n$ can be replaced by generalized wave number $q_n=n/\alpha$ . By assuming the adiabatic mode for the perturbations, we will have
\begin{align}\label{61}
  \delta D_n=\frac{\delta\rho_{Dn}}{\bar{\rho}_D+\bar{P}_D}=\frac{9\,q_n^2t^2\mathcal{R}_n^o\mathcal{T}(\kappa_n)}{10\,a^2}
\end{align}
\begin{align}\label{62}
  \psi _n=-\frac{3\,q_n^2t^2 \mathcal{R}_n^o\mathcal{T}(\kappa_n)}{5\,a^2}
\end{align}
\begin{align}\label{61}
  \delta \gamma_n=&\frac{\delta\rho_{\gamma n}}{\bar{\rho}_\gamma+\bar{P}_\gamma}=\frac{3\mathcal{R}_n^o}{5}\Bigg[\mathcal{T}(\kappa_n)(1+3R)\nonumber\\
  -&(1+R)^{-1/4}e^{-\int_{0}^{t}\Gamma\dd t}\mathcal{S}(\kappa_n)\times\nonumber\\
  &\cos\bigg(\int_{0}^{t}\frac{q_n\,\dd t}{a\sqrt{3(1+R)}}+\Delta(\kappa_n)\bigg)\Bigg]
\end{align}
\begin{align}\label{62}
  \delta u_n=&\frac{3\mathcal{R}_n^o}{5}\Bigg[-t\,\mathcal{T}(\kappa_n)
  +\frac{a}{\sqrt{3}k(1+R)^{3/4}}e^{-\int_{0}^{t}\Gamma\dd t}\mathcal{S}(\kappa_n)\times\nonumber\\
  &\sin\bigg(\int_{0}^{t}\frac{q_n\,\dd t}{a\sqrt{3(1+R)}}
  +\Delta(\kappa_n)\bigg)\Bigg]
\end{align}
where $\mathcal{T}(\kappa_n ),\; \mathcal{S}(\kappa_n ),\; \Delta(\kappa_n )$ are time-independent dimensionless function of the dimensionless rescaled wave number $\kappa_n=\frac{\sqrt{2}\,q_n}{a_{EQ}H_{EQ}}$, called transfer functions. A detail expression for these functions could be found in \cite{r20}. Also $R=\frac{3\,\overline{\rho}_B}{4\,\overline{\rho}_\gamma}$ and $\Gamma(t)$ is the acoustic damping rate.
Also, the field equations govern the scalar mode perturbations in synchronous gauge are \cite{r15}
\begin{subequations}\label{63}
  \begin{align}\label{63-a}
  2KA-&3\,a\dot{a}\dot{A}-\frac{1}{2}a\dot{a}\nabla^2\dot{B}-\frac{1}{2}a^2\ddot{A}+\frac{1}{2}\nabla^2A\nonumber\\
  =&4\pi Ga^2(-\delta\rho+\delta P+\nabla^2\Pi^S)
\end{align}
\begin{align}\label{63-b}
  -3\,a\dot{a}\dot{B}-a^2\ddot{B}+A=-16\pi Ga^2\Pi^S
\end{align}
\begin{align}\label{63-c}
  3\frac{\dot{a}}{a}\dot{A}+&\frac{\dot{a}}{a}\nabla^2 \dot{B}+\frac{3}{2}\ddot{A}+\frac{1}{2}\nabla^2\ddot{B}\nonumber\\
  =&-4\pi G(\delta\rho+3\delta P+\nabla^2\Pi^S)
\end{align}
\end{subequations}
that gives the simple relation
\begin{subequations}\label{66}
  \begin{align}\label{66-a}
  \nabla^2A+3KA=-8\pi Ga^2\delta\rho+2Ha^2\psi
\end{align}
where
\begin{align}\label{66-b}
  \psi=\frac{1}{2}(3\dot{A}+\nabla^2\dot{B})
\end{align}
\end{subequations}
Now we make the approximation that the gravitational field perturbations at last scattering are dominated by perturbations in dark matter density. So we can write above equation in terms of Fourier transform as
\begin{align}\label{68}
 -q_n^2A_n+3KA_n=-8\pi Ga^2\,\bar{\rho}_D\,\delta D_n+2Ha^2\,\psi_n
\end{align}
In the matter dominated era, we have $a\propto t^{2/3},\;\rho_D\propto a^{-3},\;\delta D_n\propto t^{2/3}$ and $\psi_n\propto t^{-1/3}$. So the right hand side of above equation will be time-independent and we have $\dot{A}_n=0$.
Therefore using the definition for the field $\psi_n,\;\dot{B}_n$ and $\ddot{B}_n$ can be written as
\begin{align}\label{69}
 \dot{B}_n=-\frac{2}{q_n^2}\psi_n\qquad\ddot{B}_n=\frac{2}{3\,t\,q_n^2}\psi_n
\end{align}
Also, recalling $\overline{\rho}_\gamma\propto T^4$ for unperturbed photon energy density, we find $\frac{\Delta T}{\overline{T}}=\frac{\delta \rho_\gamma}{4\overline{\rho}_\gamma}=\frac{1}{3}\delta_\gamma$.
Using above equations into the Eqs.(\ref{49b}),(\ref{49c}) gives the form factor $F_n$ and $G_n$ as
\begin{align}\label{70}
  F_n=&-\frac{1}{2}a^2(t_L)\ddot{B}(t_L)-\frac{1}{2}a(t_L)\dot{a}(t_L)\dot{B}_n(t_L)+\frac{\delta T_n(t_L)}{\overline{T}(t_L)}\nonumber\\
  =&\frac{1}{3}\delta_\gamma(t_L)-\frac{a^2(t_L)}{3t_L\,q_n^2}\psi_n(t_L)\nonumber\\
  =&\frac{\mathcal{R}_n^o}{5}\Bigg[3\mathcal{T}(\kappa_n)R_L\nonumber\\
  -&(1+R_L)^{-1/4}e^{-\int_{0}^{t_L}\Gamma\dd t}\mathcal{S}(\kappa_n)\times\nonumber\\
  &\cos\bigg(\int_{0}^{t_L}\frac{q_n\,\dd t}{a\sqrt{3(1+R)}}+\Delta(\kappa_n)\bigg)\Bigg]
\end{align}
\begin{align}\label{71}
  G_n=&-\left(\frac{1}{2}a(t_L)\dot{B}(t_L)+\frac{\delta u_n(t_L)}{a(t_L)}\right)\nonumber\\
  =&-\left(-\frac{a(t_L)}{q_n^2}\psi_n(t_L)+\frac{\delta u_n(t_L)}{a(t_L)}\right)\nonumber\\
  =&-\frac{\sqrt{3}\mathcal{R}_n^o}{5q_n(1+R_L)^{3/4}} e^{-\int_{0}^{t_L}\Gamma\dd t}\mathcal{S}(\kappa_n)\times\nonumber\\
  &\sin\bigg(\int_{0}^{t_L}\frac{q_n\,\dd t}{a\sqrt{3(1+R)}}+\Delta(\kappa_n)\bigg)
\end{align}

In order to recover the sudden transition from opacity to transparency assumption, we multiply $\sin$ and $\cos$ functions appeared in the form factors with a Gaussian probability function and average them in time \cite{r20}. The whole effect of this averaging is to introduce an additional damping factor $\exp(-\omega_L^2\sigma_t^2/2)$ which can be added to the acoustic damping factor
\begin{align}\label{72}
 \int_{0}^{t_L}\Gamma \dd t+\frac{\omega_L^2\sigma_t^2}{2}=\left(\frac{q_nd_D}{a_L}\right)^2
\end{align}
and introduce a new parameter $d_D$ called “damping length” as
\begin{subequations}\label{73}
  \begin{align}\label{73-a}
  d_D^2=d_{Silk}^2+d_{Landau}^2
\end{align}
where
\begin{align}\label{73-b}
  d_{Silk}^2=a^2_L\int_{0}^{t_L}\frac{t_\gamma\, c^2\, \dd t}{6a^2(1+R)}\left[\frac{16}{15}+\frac{R^2}{1+R}\right]
\end{align}
\begin{align}\label{73-c}
  d_{Landau}^2=\frac{\sigma_t^2}{6(1+R_L)}
\end{align}
\end{subequations}
We also introduce $d_H$ and $d_T$ as
\begin{align}\label{74}
  d_H=a_L\int_{0}^{t_L}\frac{c\, \dd t}{a\sqrt{3(1+R)}}
\end{align}
\begin{align}\label{75}
  d_T=\frac{0.0177}{\Omega_M h^2}\qquad \kappa_n=\frac{q_nd_T}{a_L}
\end{align}

From the theory of inflation, we know $\mathcal{R}^o_n\propto q_n^{-3/2}$ \cite{r22}. Following the flat universe parameterizations for this quantity, we have \cite{r20}
\begin{align}\label{76}
  |\mathcal{R}^o_n|^2=N^2\,q_n^{-3}\left(\frac{q_n/a_0}{k_\mathcal{R}}\right)^{n_s-1}
\end{align}
Also, the photon scattering by the free electrons produced by reionization process, can be shown simply as an overall damping factor $\exp(-2\tau_{reion})$ multiplied to the multipole coefficient \cite{r20}.\\
Putting all above equations, the quantity quoted as scalar contribution to the multipole coefficient in the spatially closed universe is
\begin{align}\label{79}
  &\frac{\ell(\ell+1)}{2\pi}C^S_{TT,\ell}=\nonumber\\
  &\frac{4\pi\,T_0^2\,N^2\exp(-2\tau_{reion})}{25}\times\nonumber\\
  &\sum_{n>\frac{\ell}{\sin\chi_L}}\frac{\ell\left(\frac{n/a_0}{\alpha\,k_\mathcal{R}}\right)^{n_s-1}}{n^2\,\sin^2\chi_L(\frac{n^2}{\ell^2}-\frac{1}{\sin^2\chi_L})^{\frac{1}{2}}}\Bigg[\bigg(3\,\mathcal{T}\left(\frac{q_nd_T}{a_L}\right)R_L\nonumber\\
  -&(1+R_L)^{-\frac{1}{4}}e^{-\left(\frac{q_nd_D}{a_L}\right)^2}\mathcal{S}\left(\frac{q_nd_T}{a_L}\right)\times\nonumber\\
  &\cos\left(\frac{q_nd_H}{a_L}+\Delta\left(\frac{q_nd_T}{a_L}\right)\right)\bigg)^2\nonumber\\
  +&\frac{3\left(1-\frac{\ell^2}{n^2\,\sin^2\chi_L}\right)}{(1+R_L)^{\frac{3}{2}}}e^{-2\left(\frac{q_nd_D}{a_L}\right)^2}\mathcal{S}^2\left(\frac{q_nd_T}{a_L}\right)\times\nonumber\\
  &\sin^2\left(\frac{q_nd_H}{a_L}+\Delta\left(\frac{q_nd_T}{a_L}\right)\right)\Bigg]
\end{align}
where
\begin{align}\label{84}
 \sin\chi_L=\sin\bigg[\frac{c}{\alpha\, a_0\,H_0}\int_{\frac{1}{1+z_L}}^{1}\frac{\dd x}{x^2\sqrt{\Omega_{s}}}\bigg],
\end{align}
while
\begin{align*}
 \Omega_{s}=\Omega_\Lambda+\Omega_Kx^{-2}+\Omega_Mx^{-3}+\Omega_Rx^{-4}
\end{align*}
and
\begin{align}\label{85}
 \Omega_K=\frac{-c^2}{\alpha^2\,a_0^2\,H_0^2}
\end{align}
Also, for the angular diameter distance of the surface of last scattering in spatially closed universe, we have
\begin{align}\label{120}
 d_A=a_Lr_L=\frac{\alpha}{(1+z_L)}\sin\chi_L
\end{align}
\section{TT Power spectrum; comparison with numerical result and observation}\label{506}
At this section, we plot the scalar mode TT power spectrum derived from the previous section using the latest cosmological parameters and compare it with CAMB code result as a numerical calculation.
In order to see how well all applied approximations work in practice, we shall calculate $C_{TT,\ell}^S$ for a realistic set of values for cosmological parameters extracted from a fit to data from latest observation “Planck 2015”. The cosmological parameters of this set are (TT+Lensing): \cite{r23}
\begin{align*}
  \Omega_Mh^2=0.1416\qquad\Omega_Bh^2=0.02225
\end{align*}
\begin{align*}
  h=0.6781\qquad\Omega_k=1-\Omega_M-\Omega_\Lambda-\Omega_R=-0.005
\end{align*}
And	
\begin{align*}
  n_s=0.9677\qquad k_{\mathcal{R}}=0.05\,Mpc^{-1}
\end{align*}
\begin{align*}
 N^2=\frac{A_s}{4\pi}=1.702\times10^{-10}\qquad \exp(-2\tau_{reion})=0.8763
\end{align*}
We take $T_0=2.725 K$ , Which yields $\Omega_\gamma h^2=2.47\times10^{-5}$ , and by assuming three flavors of massless neutrino we have $\Omega_R h^2=4.15\times10^{-5}$. We will also adopt the parameters describing recombination as
\begin{align*}
  1+z_L=1090\quad\sigma_t=262\,K\quad t_L=370,000\,yrs
\end{align*}
By using above values we find
\begin{align*}
  R_0=\frac{3\Omega_B}{4\Omega_\gamma}=675.99
\end{align*}
\begin{align*}
  R_L=\frac{3\Omega_B}{4\Omega_\gamma(1+z_L)}=0.6201
\end{align*}
\begin{align*}
  R_{EQ}=\frac{3\Omega_B\Omega_R}{4\Omega_\gamma\Omega_M}=0.1979
\end{align*}
and Eqs.(\ref{73}),(\ref{74}),(\ref{75}),(\ref{120}) gives
\begin{align*}
  d_A=12.65\,Mpc\qquad d_T=0.1251\,Mpc
\end{align*}
\begin{align*}
  d_H=0.1331\,Mpc\qquad d_D=0.008056\,Mpc
\end{align*}
while the overall factor multiplying the integral is
\begin{align}\label{121}
  &\frac{4\pi\,T_0^2\,N^2\exp(-2\tau_{reion})}{25}=556.77\,\mu K^2
\end{align}

\begin{figure}
\includegraphics[width=0.45\textwidth]{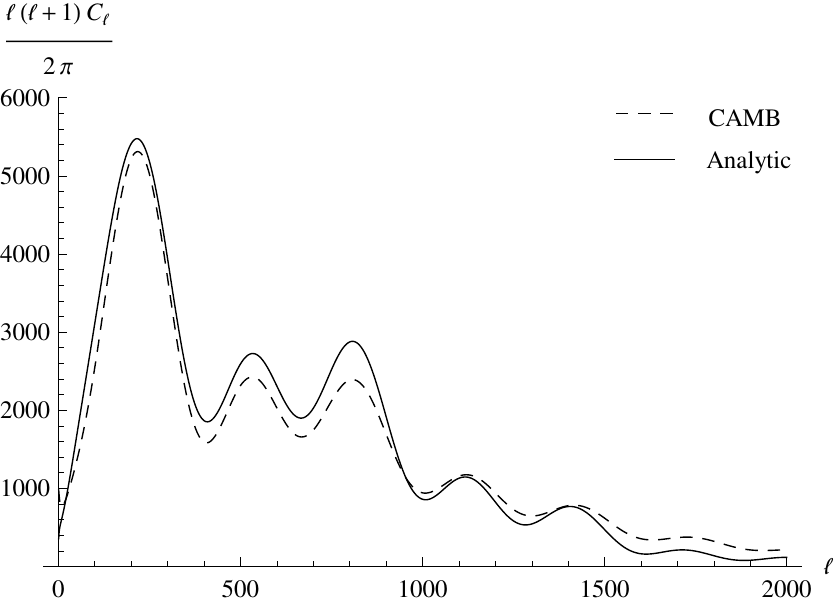}
\caption{The scalar multipole coefficient $\ell(\ell+1)C^S_{TT,\ell}/2\pi$ in square micreoKelvin, vs. $\ell$, for a closed universe with $\Omega_K=-0.005$ in comparison with numerical result from CAMB.}
\label{An-CAMB1}
\end{figure}

\begin{figure}
\includegraphics[width=0.45\textwidth]{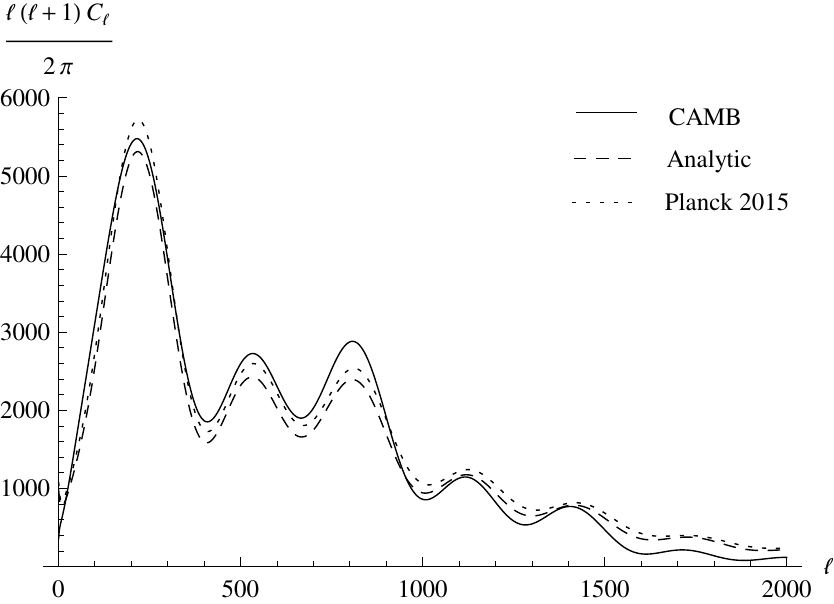}
\caption{The scalar multipole coefficient $\ell(\ell+1)C^S_{TT,\ell}/2\pi$ in square micreoKelvin, vs. $\ell$, for a closed universe with $\Omega_K=-0.005$ in comparison with numerical result from CAMB and Planck 2015 observational result.}
\label{An-CAMB-Planck}
\end{figure}

\begin{figure}
\includegraphics[width=0.45\textwidth]{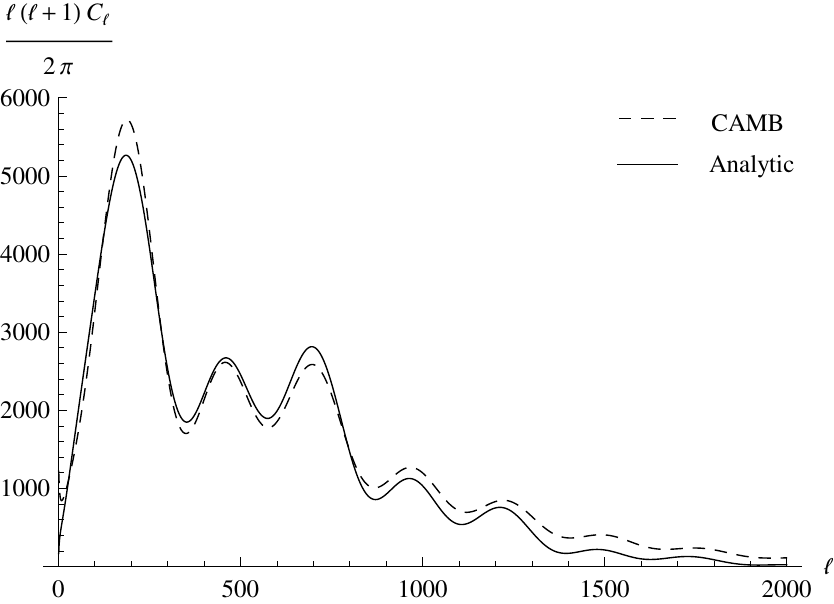}
\caption{The scalar multipole coefficient $\ell(\ell+1)C^S_{TT,\ell}/2\pi$ in square micreoKelvin, vs. $\ell$, for a closed universe with $\Omega_K=-0.1$ in comparison with numerical result from CAMB.}
\label{An-CAMB2}
\end{figure}

Finally, based on Eq.(\ref{79}), above parameters and sum over $n$ until 10000, the scalar multipole coefficient power spectrum in closed universe with $\Omega_k=-0.005$ in comparison with numerical calculation based on the CAMB code \cite{r23} is shown in figure \ref{An-CAMB1}.
This figure shows that the overall profile of the analytic spectra agrees well with the numerical result for almost all $\ell$. The peak positions are in very good agreement with numerical result while the peak heights agree with numerical curve to within $10\%$ due to the approximations have been considered for this derivation (sudden transition from opacity to transparency and take the evolution of perturbations hydrodynamically) \cite{r19}.\\
Just for comparison, the observational result from Planck is also given in figure \ref{An-CAMB-Planck}.\\
In order to see how this formalism works for the bigger amount of $\Omega_k$, we consider the purely abstract amount of $\Omega_k=-0.1$ (while other parameters remain unchanged) and plot the result in comparison with numerical one in figure \ref{An-CAMB2}. As it is clear from this figure, the overall result is same as small amount of $\Omega_k$, while the discrepancy for the peak heights between numerical and analytic curves has shifted to larger amount of $\ell$.\\

\begin{figure}
\includegraphics[width=0.45\textwidth]{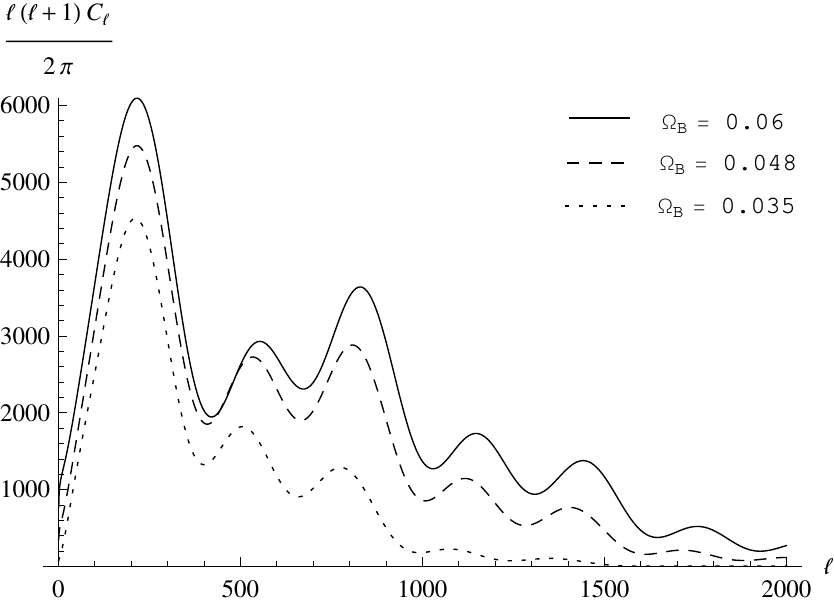}
\caption{Baryon density $\Omega_B$ dependency of the scalar multipole coefficient $\ell(\ell+1)C^S_{TT,\ell}/2\pi$ in square micreoKelvin, vs. $\ell$, for a closed universe with $\Omega_K=-0.005$}
\label{An-OB}
\end{figure}

\begin{figure}
\includegraphics[width=0.45\textwidth]{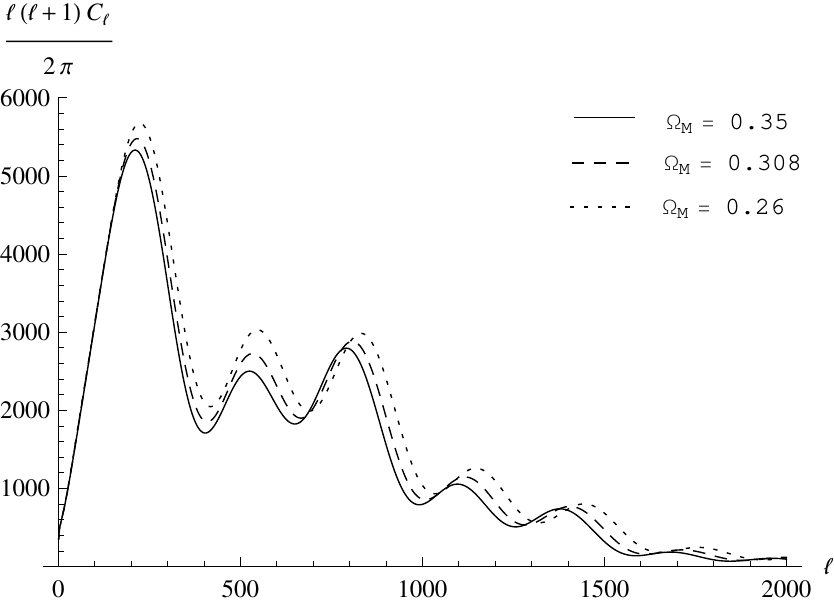}
\caption{Matter density $\Omega_M$ dependency of the scalar multipole coefficient $\ell(\ell+1)C^S_{TT,\ell}/2\pi$ in square micreoKelvin, vs. $\ell$, for a closed universe with $\Omega_K=-0.005$}
\label{An-OM}
\end{figure}

\begin{figure}
\includegraphics[width=0.45\textwidth]{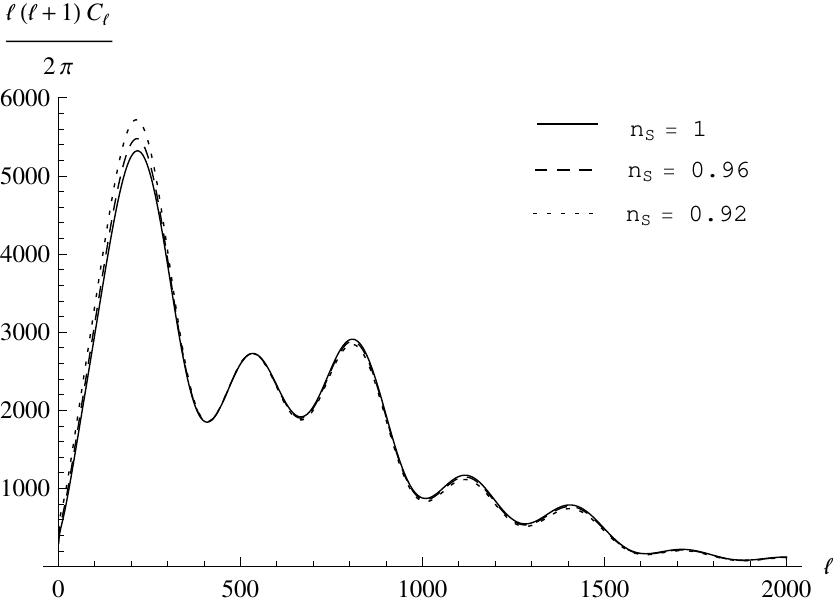}
\caption{Primordial spectral index $n_S$ dependency of the scalar multipole coefficient $\ell(\ell+1)C^S_{TT,\ell}/2\pi$ in square micreoKelvin, vs. $\ell$, for a closed universe with $\Omega_K=-0.005$}
\label{An-n}
\end{figure}

\begin{figure}
\includegraphics[width=0.45\textwidth]{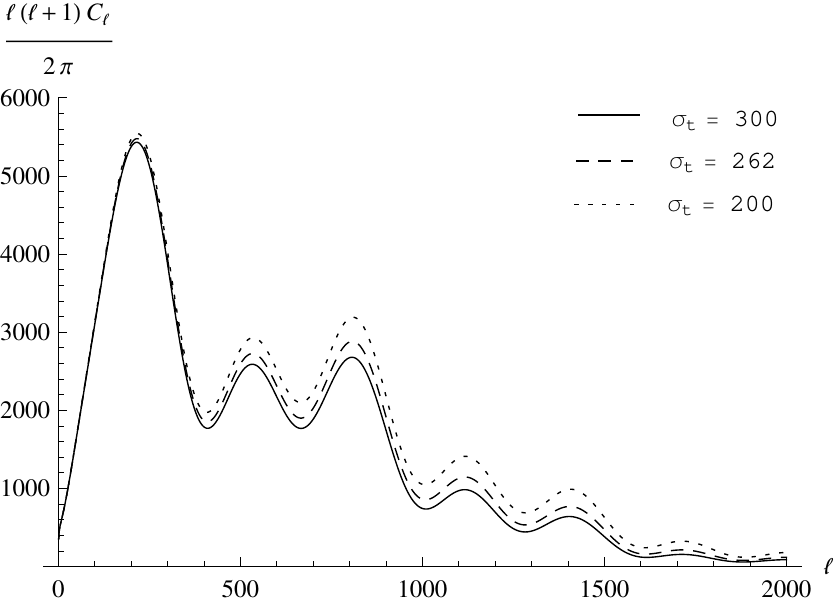}
\caption{Recombination width $\sigma_t$ dependency of the scalar multipole coefficient $\ell(\ell+1)C^S_{TT,\ell}/2\pi$ in square micreoKelvin, vs. $\ell$, for a closed universe with $\Omega_K=-0.005$}
\label{An-Sigma}
\end{figure}

\begin{figure}
\includegraphics[width=0.45\textwidth]{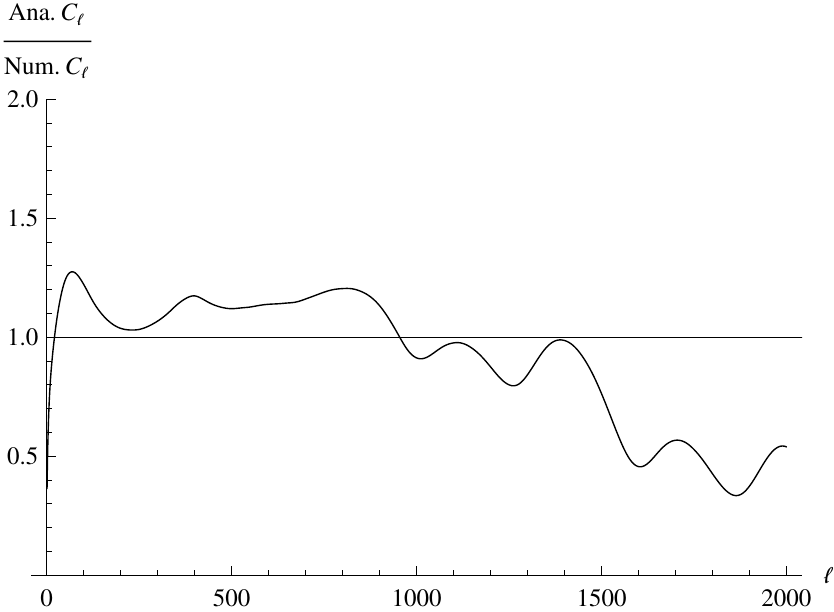}
\caption{The ratio of the analytic spectra to numerical spectra for a closed universe with $\Omega_K=-0.005$}
\label{Div-An-CAMB}
\end{figure}

Figure \ref{An-OB} shows the dependence of $C^S_{TT,\ell}$ upon the baryon fraction $\Omega_B$. This figure shows that a greater value of $\Omega_B$ yields higher amplitude of $C^S_{TT,\ell}$. This can be understood as follows. In Eq.(\ref{79}) the sum over $n$ receives its greatest contribution from $n\approx\frac{\ell}{\sin\chi_L}$, so the Doppler term makes a relatively small contribution to the multipole coefficients and the sum over $n$ will be dominated by the term proportional to $R_L$. A greater $\Omega_B$ correspond to a greater $R_L$ that gives the higher amplitude of $C^S_{TT,\ell}$. This figure also shows that a greater $\Omega_B$ shifts the location of peaks to larger $\ell$. This is because a greater $\Omega_B$ leads to lower sound speed $C_s$ of photon gas (hence a lower acoustic horizon distance $d_H$). so at a fixed frequency the corresponding wavelength $q_n$ suppressed. By the analytic result, this is evident from oscillating factors $\cos(\frac{q_nd_H}{a_L})$ and $\sin(\frac{q_nd_H}{a_L})$, whose peak locations are stretched to a larger wave number $q_n$ (i.e. larger $\ell$) for smaller $d_H$.\\
Figure \ref{An-OM} shows that smaller amount of $\Omega_M$ enhance slightly the amplitude of $C^S_{TT,\ell}$ through transfer function $\mathcal{S}$ among a shift to higher $\ell$ through the phase shift in oscillating terms by transfer function $\Delta$.\\
In figures \ref{An-OB} and \ref{An-OM}, the amount of $\Omega_\Lambda$ has been changed in order to keep $\Omega_K$ fixed.\\
Figure \ref{An-n} shows the dependence of $C^S_{TT,\ell}$ upon the scalar spectral index $n_s$. As is seen, a lower value of $n_s$ yields a higher amplitude of $C^S_{TT,\ell}$ which is expected from analytic expression. The effect is most obvious around the first peak.\\
Figure \ref{An-Sigma} shows that a longer recombination process (a greater $\sigma_t$) yields a lower amplitude of $C^S_{TT,\ell}$ and also more damping on smaller scales due to increasing the damping length $d_D$.\\
Figure \ref{Div-An-CAMB} shows the ratio of analytic spectrum to the numerical one that is centered around $1$ for $\ell\leq1000$, showing a reasonable agreement between the analytic and numeric analysis on large and moderate angular scales.\\

\section{Conclusion and summary}\label{507}
We have examined the CMB anisotropies in a model with the spatially closed background. By considering adiabatic mode for the perturbations and considering a sudden transition from opacity to transparency, we extracted a formula for the scalar mode temperature fluctuation in a closed universe from two different methods. Also, an \emph{analytic} formula for the multipole coefficient in a closed universe has been extracted. This gives a great insight into the problem by providing transparent information about the CMB anisotropies and it's dependence on cosmological parameters in a spatially closed background than other works at the field that gives a generally complicated formula more useful for computer calculations. This achieves by means of some approximations at the expense of lack of small accuracy. The first approximation we have made is assuming a sudden transition from opacity to transparency at a definite time $t_L$, but of course the drop takes place during some finite interval of time, The next one is neglecting the integrated Sachs-Wolfe effect that is important only for relatively small values of $\ell$, where cosmic variance intrudes on measurement of $C_{TT,\ell}^S$, the third main approximation we have made is considering the evolution of perturbation hydrodynamically and consequently neglect the anisotropy inertia. We also have made the approximation that the gravitational field perturbations at last scattering are dominated by perturbations in dark matter density.\\
The procedure introduced for extracting temperature fluctuations from the Boltzmann equation using the line of sight integral method can be employed to extract the polarization of the cosmic microwave background and relevant multipole coefficients $C_{TE,\ell}^S  ,\;C_{EE,\ell}^S$.\\
We compared the hydrodynamically extracted scalar mode temperature multipole coefficient power spectrum in the closed background with the result of CAMB as a numerical calculation and find a general agreement between the analytic and numeric analysis on large and moderate angular scales.
As the major advantage of analytic expression, $C_{TT,\ell}^S$ explicitly shows the dependencies on baryon density $\Omega_B$, matter density $\Omega_M$, curvature $\Omega_K$, primordial spectral index $n_s$, primordial power spectrum amplitude $A_s$, Optical depth $\tau_{reion}$, recombination width $\sigma_t$ and recombination time $t_L$ that are not transparent in numerical codes.

\appendix*
\section{}
We expand $J^{ij} (\textbf{x},\hat{p},t)$, the metric perturbations and $\delta u_k (\textbf{x},t)$ in terms of the eigenvalues of Laplace-Beltrami operator.
We may write the $J^{ij} (\textbf{x},\hat{p},t)$  matrix as
\begin{align*}
  J^{ij} (\textbf{x},\hat{p},t)=&\sum_{n\ell m}\alpha_{\ell m}\bigg[\frac{1}{2}\left(\Delta_{Tn}(t)-\Delta_{Pn}(t)\right)\times\nonumber\\
  &\left(\tilde{g}^{ij}-\hat{p}^i\hat{p}^j\right)\mathcal{Y}_{n\ell m}(\chi,\theta,\phi)\nonumber\\
  +&\,\Delta_{Pn}(t)k_n^{-2}q^{ij}\mathcal{Y}_{n\ell m}(\chi,\theta,\phi)\bigg]
\end{align*}
where
\begin{equation*}
  q^{ij}=\frac{(\nabla^i-\hat{p}^i\hat{p}_s\nabla^s)(\nabla^j-\hat{p}^j\hat{p}_t\nabla^t)}{\mathcal{Y}^{-1}_{n\ell m}(\chi,\theta,\phi)k_n^{-2}(\nabla^2-\hat{p}_s\hat{p}_tH^{st})\mathcal{Y}_{n\ell m}(\chi,\theta,\phi)}
\end{equation*}
The metric perturbation in scalar mode can be written as
\begin{equation*}
  a^{-2}(t)\delta g_{st}=A\,\tilde{g}_{st}+H_{st}\,B
\end{equation*}
so
\begin{equation*}
  \hat{p}^s\hat{p}^t\pdv{t}\left(a^{-2}\delta g_{st}\right)=\dot{A}+\hat{p}^s\hat{p}^t\,H_{st}\,\dot{B}
\end{equation*}
where for the perturbative quantities $A$ and $B$ we can write
\begin{align*}
 A(\textbf{x},t)&=\sum_{n\ell m}\alpha_{\ell m}A_n(t)\mathcal{Y
 }_{n\ell m}(\chi,\theta,\phi)\\
 B(\textbf{x},t)&=\sum_{n\ell m}\alpha_{\ell m}B_n(t)\mathcal{Y
 }_{n\ell m}(\chi,\theta,\phi)
\end{align*}
The plasma velocity can be expressed in terms of the velocity potential and then expand as
\begin{equation*}
 \delta u_k(\textbf{x},t)=\nabla_k\,\delta u(\textbf{x},t)=\sum_{n\ell m}\alpha_{\ell m}\delta u_n(t)\nabla_k\,\mathcal{Y}_{n\ell m}(\chi,\theta,\phi)
\end{equation*}
Also, we introduce the source functions $\varphi_n (t)$ and $\mathcal{J}_n (t)$ as
\begin{align*}
 \int\frac{\dd[2]{\hat{p}_1}}{4\pi}&\sqrt{Det\,\tilde{g}^{ij}}J^{ij}(x,p_1,t)\\
 &=\sum_{n\ell m}\alpha_{\ell m}\Big[\varphi_n(t)\tilde{g}^{ij}\mathcal{Y}_{n\ell m}(\chi,\theta,\phi)\\
 &-\frac{1}{2}\mathcal{J}_n(t)H^{ij}k_n^{-2}\mathcal{Y}_{n\ell m}(\chi,\theta,\phi)\Big]
\end{align*}
By inserting all above relations into the Eq.(\ref{26}), we have:
\begin{align}\label{101}
 &\frac{1}{2}\left(\dot{\Delta}_{Tn}(t)-\dot{\Delta}_{Pn}(t)\right)\left(\tilde{g}^{ij}-\hat{p}^i\hat{p}^j\right)\mathcal{Y}_{n\ell m}\nonumber\\
 &+\dot{\Delta}_{Pn}(t)k_n^{-2}q^{ij}\mathcal{Y}_{n\ell m}\nonumber\\
 &+\frac{\hat{p}^k}{2\,a(t)}\left(\Delta_{Tn}(t)-\Delta_{Pn}(t)\right)\left(\tilde{g}^{ij}-\hat{p}^i\hat{p}^j\right)\nabla_k\,\mathcal{Y}_{n\ell m}\nonumber\\
 &+\frac{\hat{p}^k}{a(t)}\Delta_{Pn}(t)k_n^{-2}q^{ij}\nabla_k\,\mathcal{Y}_{n\ell m}\nonumber\\
 &+\left(\dot{A}+\hat{p}^s\hat{p}^t\,H_{st}\,\dot{B}\right)\left(\tilde{g}^{ij}-\hat{p}^i\hat{p}^j\right)\mathcal{Y}_{n\ell m}\nonumber\\
 &-\frac{\hat{p}^k\hat{p}^s\hat{p}^t}{a(t)}\partial_k\,\tilde{g}_{st}\left(\Delta_{Tn}(t)-\Delta_{Pn}(t)\right)\left(\tilde{g}^{ij}-\hat{p}^i\hat{p}^j\right)\mathcal{Y}_{n\ell m}\nonumber\\
 &-2\,\frac{\hat{p}^k\hat{p}^m\hat{p}^n}{a(t)}\partial_k\,\tilde{g}_{mn}\Delta_{Pn}(t)k_n^{-2}q^{ij}\mathcal{Y}_{n\ell m}\nonumber\\
 &=-\frac{1}{2}\omega_c(t)\left(\Delta_{Tn}(t)-\Delta_{Pn}(t)\right)\left(\tilde{g}^{ij}-\hat{p}^i\hat{p}^j\right)\mathcal{Y}_{n\ell m}\nonumber\\
 &-\omega_c(t)\Delta_{Pn}(t)k_n^{-2}q^{ij}\mathcal{Y}_{n\ell m}\nonumber\\
 &+\frac{3\,\omega_c(t)}{2}\Big[\varphi_n(t)\tilde{g}^{ij}\mathcal{Y}_{n\ell m}-\frac{1}{2}\mathcal{J}_n(t)H^{ij}k_n^{-2}\mathcal{Y}_{n\ell m}\nonumber\\
 &-\hat{p}^i\hat{p}_r\big(\varphi_n(t)\tilde{g}^{jr}\mathcal{Y}_{n\ell m}-\frac{1}{2}\mathcal{J}_n(t)H^{jr}k_n^{-2}\mathcal{Y}_{n\ell m}\big)\nonumber\\
 &-\hat{p}^j\hat{p}_r\big(\varphi_n(t)\tilde{g}^{ir}\mathcal{Y}_{n\ell m}-\frac{1}{2}\mathcal{J}_n(t)H^{ir}k_n^{-2}\mathcal{Y}_{n\ell m}\big)\nonumber\\
 &+\hat{p}^i\hat{p}^j\hat{p}_s\hat{p}_t\big(\varphi_n(t)\tilde{g}^{st}\mathcal{Y}_{n\ell m}-\frac{1}{2}\mathcal{J}_n(t)H^{st}k_n^{-2}\mathcal{Y}_{n\ell m}\big)\Big]\nonumber\\
 &+\frac{2\,\omega_c(t)}{a(t)}\left(\tilde{g}^{ij}-\hat{p}^i\hat{p}^j\right)\hat{p}^k\delta u_n(t)\nabla_k\,\mathcal{Y}_{n\ell m}
\end{align}
The statement inside the bracket could be simplified as
\begin{align}\label{102}
 &\varphi_n(t)\tilde{g}^{ij}\mathcal{Y}_{n\ell m}-\frac{1}{2}\mathcal{J}_n(t)H^{ij}k_n^{-2}\mathcal{Y}_{n\ell m}\nonumber\\
 &-\hat{p}^i\hat{p}^j\varphi_n(t)\mathcal{Y}_{n\ell m}+\frac{1}{2}\mathcal{J}_n(t)\hat{p}^i\hat{p}_r H^{jr}k_n^{-2}\mathcal{Y}_{n\ell m}\nonumber\\
 &-\hat{p}^j\hat{p}^i\varphi_n(t)\mathcal{Y}_{n\ell m}-\frac{1}{2}\mathcal{J}_n(t)\hat{p}^j\hat{p}_r H^{ir}k_n^{-2}\mathcal{Y}_{n\ell m}\nonumber\\
 &+\hat{p}^i\hat{p}^j\varphi_n(t)\mathcal{Y}_{n\ell m}-\frac{1}{2}\mathcal{J}_n(t)\hat{p}^i\hat{p}^j\hat{p}_s\hat{p}_t H^{st}k_n^{-2}\mathcal{Y}_{n\ell m}\nonumber\\
 &=\varphi_n(t)\left(\tilde{g}^{ij}-\hat{p}^j\hat{p}^i\right)\mathcal{Y}_{n\ell m}\nonumber\\
 &-\frac{1}{2}\mathcal{J}_n(t)\Big(H^{ij}+\hat{p}^i\hat{p}_r H^{jr}-\hat{p}^j\hat{p}_r H^{ir}\nonumber\\
 &-\hat{p}^i\hat{p}^j\hat{p}_s\hat{p}_t H^{st}\Big)k_n^{-2}\mathcal{Y}_{n\ell m}\nonumber\\
 &=\varphi_n(t)\left(\tilde{g}^{ij}-\hat{p}^j\hat{p}^i\right)\mathcal{Y}_{n\ell m}\nonumber\\
 &-\frac{1}{2}\mathcal{J}_n(t)(\nabla^i-\hat{p}^i\hat{p}_s\nabla^s)(\nabla^j-\hat{p}^j\hat{p}_t\nabla^t)k_n^{-2}\mathcal{Y}_{n\ell m}\nonumber\\
 &=\varphi_n(t)\left(\tilde{g}^{ij}-\hat{p}^j\hat{p}^i\right)\mathcal{Y}_{n\ell m}\nonumber\\
 &-\frac{1}{2}\mathcal{J}_n(t)k_n^{-2}(\nabla^2-\hat{p}_s\hat{p}_tH^{st})k_n^{-2}\mathcal{Y}_{n\ell m}q^{ij}
\end{align}
so the Eq.(\ref{101}) can be written as
\begin{align}\label{103}
 &\frac{1}{2}\left(\dot{\Delta}_{Tn}(t)-\dot{\Delta}_{Pn}(t)\right)\left(\tilde{g}^{ij}-\hat{p}^i\hat{p}^j\right)\mathcal{Y}_{n\ell m}\nonumber\\
 &+\dot{\Delta}_{Pn}(t)k_n^{-2}q^{ij}\mathcal{Y}_{n\ell m}\nonumber\\
 &+\frac{\hat{p}^k}{2\,a(t)}\left(\Delta_{Tn}(t)-\Delta_{Pn}(t)\right)\left(\tilde{g}^{ij}-\hat{p}^i\hat{p}^j\right)\nabla_k\,\mathcal{Y}_{n\ell m}\nonumber\\
 &+\frac{\hat{p}^k}{a(t)}\Delta_{Pn}(t)k_n^{-2}q^{ij}\nabla_k\,\mathcal{Y}_{n\ell m}\nonumber\\
 &+\left(\dot{A}+\hat{p}^s\hat{p}^t\,H_{st}\,\dot{B}\right)\left(\tilde{g}^{ij}-\hat{p}^i\hat{p}^j\right)\mathcal{Y}_{n\ell m}\nonumber\\
 &-\frac{\hat{p}^k\hat{p}^s\hat{p}^t}{a(t)}\partial_k\,\tilde{g}_{st}\left(\Delta_{Tn}(t)-\Delta_{Pn}(t)\right)\left(\tilde{g}^{ij}-\hat{p}^i\hat{p}^j\right)\mathcal{Y}_{n\ell m}\nonumber\\
 &-2\,\frac{\hat{p}^k\hat{p}^m\hat{p}^n}{a(t)}\partial_k\,\tilde{g}_{mn}\Delta_{Pn}(t)k_n^{-2}q^{ij}\mathcal{Y}_{n\ell m}\nonumber\\
 &=-\frac{1}{2}\omega_c(t)\left(\Delta_{Tn}(t)-\Delta_{Pn}(t)\right)\left(\tilde{g}^{ij}-\hat{p}^i\hat{p}^j\right)\mathcal{Y}_{n\ell m}\nonumber\\
 &-\omega_c(t)\Delta_{Pn}(t)k_n^{-2}q^{ij}\mathcal{Y}_{n\ell m}\nonumber\\
 &+\frac{3\,\omega_c(t)}{2}\Big[\varphi_n(t)\left(\tilde{g}^{ij}-\hat{p}^j\hat{p}^i\right)\mathcal{Y}_{n\ell m}\nonumber\\
 &-\frac{1}{2}\mathcal{J}_n(t)k_n^{-2}(\nabla^2-\hat{p}_s\hat{p}_tH^{st})k_n^{-2}\mathcal{Y}_{n\ell m}q^{ij}\Big]\nonumber\\
 &+\frac{2\,\omega_c(t)}{a(t)}\left(\tilde{g}^{ij}-\hat{p}^i\hat{p}^j\right)\hat{p}^k\delta u_n(t)\nabla_k\,\mathcal{Y}_{n\ell m}
\end{align}
By considering the terms proportional to $\left(\tilde{g}^{ij}-\hat{p}^i\hat{p}^j\right)$ and $q^{ij}$, above equation could be decomposed into two coupled Boltzmann equations as
\begin{align}\label{104}
 &\frac{1}{2}\left(\dot{\Delta}_{Tn}(t)-\dot{\Delta}_{Pn}(t)\right)\mathcal{Y}_{n\ell m}\nonumber\\
 &+\frac{\hat{p}^k}{2\,a(t)}\left(\Delta_{Tn}(t)-\Delta_{Pn}(t)\right)\nabla_k\,\mathcal{Y}_{n\ell m}\nonumber\\
 &+\left(\dot{A}+\hat{p}^s\hat{p}^t\,H_{st}\,\dot{B}\right)\mathcal{Y}_{n\ell m}\nonumber\\
 &-\frac{\hat{p}^k\hat{p}^s\hat{p}^t}{a(t)}\partial_k\,\tilde{g}_{st}\left(\Delta_{Tn}(t)-\Delta_{Pn}(t)\right)\mathcal{Y}_{n\ell m}\nonumber\\
 &=-\frac{1}{2}\omega_c(t)\left(\Delta_{Tn}(t)-\Delta_{Pn}(t)\right)\mathcal{Y}_{n\ell m}\nonumber\\
 &+\frac{3\,\omega_c(t)}{2}\varphi_n(t)\mathcal{Y}_{n\ell m}+\frac{2\,\omega_c(t)}{a(t)}\hat{p}^k\delta u_n(t)\nabla_k\,\mathcal{Y}_{n\ell m}
\end{align}
and
\begin{align}\label{105}
 &+\dot{\Delta}_{Pn}(t)\mathcal{Y}_{n\ell m}+\frac{\hat{p}^k}{a(t)}\Delta_{Pn}(t)\nabla_k\,\mathcal{Y}_{n\ell m}\nonumber\\
 &-2\,\frac{\hat{p}^k\hat{p}^s\hat{p}^t}{a(t)}\partial_k\,\tilde{g}_{st}\Delta_{Pn}(t)\mathcal{Y}_{n\ell m}\nonumber\\
 &=-\omega_c(t)\Delta_{Pn}(t)\mathcal{Y}_{n\ell m}\nonumber\\
 &-\frac{3\,\omega_c(t)}{4}\mathcal{J}_n(t)\left(\nabla^2-\hat{p}_s\hat{p}_tH^{st}\right)k_n^{-2}\mathcal{Y}_{n\ell m}
\end{align}
In this particular coordinate $(\chi,\theta,\phi)$, the momentum $\hat{p}$ for the photon coming from direction $\hat{n}$ will be $ \hat{p}=-\hat{n}=(-1,0 ,0)=-e_\chi$.\\
So, we can write
\begin{align*}
\hat{p}^k\nabla_k\mathcal{Y}_{n\ell m}(\chi,\theta,\varphi)&=-\nabla_\chi\mathcal{Y}_{n\ell m}(\chi,\theta,\varphi)\\
 &=-Y_{\ell m} (\theta,\varphi)\dv{\chi}\Pi_{n\ell}(\chi)
\end{align*}
\begin{align*}
 \hat{p}^s\hat{p}^tH_{st}\mathcal{Y}_{n\ell m}(\chi,\theta,\varphi)&=\nabla_\chi\nabla_\chi\mathcal{Y}_{n\ell m}(\chi,\theta,\varphi)\\
 &=Y_{\ell m} (\theta,\varphi)\dv[2]{\chi}\Pi_{n\ell} (\chi)
\end{align*}
and
\begin{align*}
\hat{p}^k\hat{p}^m\hat{p}^n\partial_k\tilde{g}_{mn}=-\partial_\chi \tilde{g}_{\chi\chi}=0
\end{align*}
Also, it is noted that
\begin{align*}
  &\nabla^2\,\mathcal{Y}_{n\ell m} (\chi,\theta,\varphi) =  -k_n^2\, \mathcal{Y}_{n\ell m} (\chi,\theta,\varphi)
\end{align*}
By using Eq.(\ref{105}) into Eq.(\ref{104}) and also above relations, the Eqs.(\ref{104}),(\ref{105}) can be written as
\begin{align}\label{106}
  &\dot{\Delta}_{Tn}(t)\Pi_{n \ell}(\chi)-\Delta_{Tn}(t)\frac{1}{a(t)}\dv{\chi}\Pi_{n \ell}(\chi)\nonumber\\
  &+2\dot{A}_n(t)\Pi_{n \ell}(\chi)+2\dot{B}_n(t)\dv[2]{\chi}\Pi_{n \ell}(\chi)\nonumber\\
  &=-\omega_c(t)\Delta_{Tn}(t)\Pi_{n\ell}(\chi)+3\,\omega_c(t)\varphi_n(t)\Pi_{n\ell}(\chi)\nonumber\\
  &-\frac{4\,\omega_c(t)}{a(t)}\delta u_n(t)\dv{\chi}\Pi_{n\ell}(\chi)\nonumber\\
  &+\frac{3}{4}\omega_c(t)\mathcal{J}_n(t)\left(1+k_n^{-2}\dv[2]{\chi}\right)\Pi_{n\ell}(\chi)
\end{align}
and
\begin{align}\label{107}
  &\dot{\Delta}_{Pn}(t)\Pi_{n \ell}(\chi)-\Delta_{Pn}(t)\frac{1}{a(t)}\dv{\chi}\Pi_{n \ell}(\chi)\nonumber\\
  &=-\omega_c(t)\Delta_{Pn}(t)\Pi_{n\ell}(\chi)\nonumber\\
  &+\frac{3}{4}\omega_c(t)\mathcal{J}_n(t)\left(1+k_n^{-2}\dv[2]{\chi}\right)\Pi_{n\ell}(\chi).
\end{align}

\end{document}